\documentclass[%
    reprint,
    amsmath,
    amssymb,
    aps,
    pra,
    longbibliography,
    showkeys,
    superscriptaddress,
]{revtex4-2}

\usepackage{graphicx}
\usepackage{dcolumn}
\usepackage{hyperref} 

\usepackage{xcolor} 
\usepackage[per-mode=symbol]{siunitx} 

\DeclareSIUnit\angstrom{\text{\AA}}         
\DeclareSIUnit\elementarycharge{\text{e}}   
\DeclareSIUnit\atm{\text{atm}}              
\DeclareSIUnit\amu{\text{amu}}              

\hypersetup{colorlinks=true, 
    linkcolor={blue},
    citecolor={blue}, 
    urlcolor={blue}
}

\begin{document}

\title{Lattice Excitations with Finite Polarization and Magnetization}

\author{Mike Pols}
\email{mike.pols@mat.ethz.ch}
\altaffiliation{%
    Present address: Materials Theory, ETH Zurich, CH-8093 Z\"urich, Switzerland
}%
\affiliation{%
    Department of Applied Physics and Science Education, Eindhoven University of Technology, 5612 AP Eindhoven, Netherlands
}%

\author{Carl P. Romao}
\email{carl.romao@cvut.cz}
\affiliation{%
    Department of Materials, Faculty of Nuclear Sciences and Physical Engineering, Czech Technical University in Prague, 
    Prague 120 00, Czech Republic
}%

\author{Dominik M. Juraschek}
\email{d.m.juraschek@tue.nl}
\affiliation{%
    Department of Applied Physics and Science Education, Eindhoven University of Technology, 5612 AP Eindhoven, Netherlands
}%

\date{\today}

\begin{abstract}

Ferrons are a type of quasiparticle corresponding to elementary excitations of the ferroelectric order. Analogously to how magnons modulate and transport magnetization, ferrons modulate and transport electric polarization. Here, we introduce \textit{multiferrons} as elementary excitations with both electric and magnetic character. Multiferrons lead to a tilt and elliptical precession of the polarization and at the same time create a magnetization through the mechanism of dynamical multiferroicity. Using first-principles calculations for LiNbO$_3$, we show that the electric polarization of multiferrons is perpendicular to the equilibrium ferroelectric polarization, whereas the magnetization is parallel to it. Our calculations further demonstrate that multiferrons carry net electric and magnetic quadrupole and octupole moments, which we term multipolons. These multipolons could couple to internal multipolar degrees of freedom, for example in altermagnets, or to external probes such as neutrons, leading to potentially experimentally observable phenomena following coherent or thermal excitation of multiferrons.

\end{abstract}


\maketitle


\section{Introduction}
\begin{figure*}[t]
    \includegraphics[width=\textwidth]{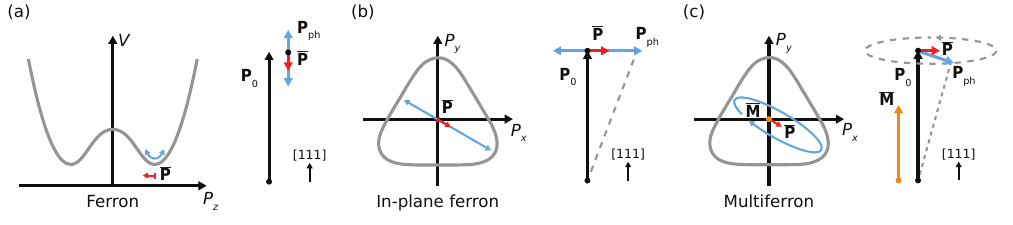}
    \caption{Ferrons in LiNbO$_3$. (a) Excitation of $A_1$ modes leads to anharmonic oscillations of the electric polarization $\mathbf{P}_\mathrm{ph}$, generating ferrons with a net polarization $\overline{\mathbf{P}}$, reducing the magnitude of the ferroelectric polarization $\mathbf{P}_0$. (b) Linear excitation of anharmonic $E$ modes produces in-plane ferrons with a net polarization perpendicular to the ferroelectric polarization, $\overline{\mathbf{P}}\perp\mathbf{P}_0$, leading to a tilting and increase of the total polarization (dashed line). (c) Elliptical excitation of anharmonic $E$ modes yields multiferrons, in which the total polarization precesses, leading to a net in-plane polarization and a net out-of-plane magnetization $\overline{\mathbf{M}}$. Polarization dynamics are shown in blue, net ferron polarization in red, and magnetization in orange.}
    \label{fig-1:overview}
\end{figure*}

Quasiparticles are fundamental excitations of ordered electronic and structural phases in solids. A prototypical example are magnons, excitations of magnetic order that lead to spin precession and a change in magnetization or Ne\'{e}l vector along the equilibrium orientation~\cite{chumakMagnonSpintronics2015, rezendeIntroductionAntiferromagneticMagnons2019}. Recently, ferrons have emerged analogously as fundamental excitations of ferroelectric order that carry a net polarization opposing the equilibrium ferroelectric polarization, as illustrated in Fig.~\ref{fig-1:overview}a~\cite{tangExcitationsFerroelectricOrder2022, bauerMagnonicsVsFerronics2022, bauerPolarizationTransportFerroelectrics2023, zhouSurfaceFerronExcitations2023, tangElectricAnalogMagnons2024, shenObservationFerronTransport2025, choeObservationCoherentFerrons2025, morozovskaFlexocouplinginducedPhononsFerrons2025, zhaoRoleFerronsHeat2025}. Ferrons can be described as electric dipole-carrying phonons existing in the intrinsically anharmonic potential energy landscape of ferroelectrics, leading to new forms of polarization transport and control \cite{bauerPerspectiveFerrons2023}.

Here, we introduce a new type of quasiparticle called the \textit{multiferron}, a fundamental excitation of nonmagnetic ferroelectrics that carries both an electric polarization and a magnetization. Multiferrons can be described as elliptically polarized in-plane ferrons that carry a net polarization perpendicular to the equilibrium ferroelectric polarization (Fig.~\ref{fig-1:overview}b) and carry a magnetization parallel to it (Fig.~\ref{fig-1:overview}c). This leads to a tilt and precession of the total electric polarization in the system. We perform first-principles calculations for multiferrons in the prototypical ferroelectric lithium niobate (LiNbO$_3$), for which we compute the quantized polarization and magnetization, as well as the dynamical response to the coherent excitation by an ultrashort terahertz pulse. We find that in addition to net electric polarization and magnetization, multiferrons also carry net quadrupole and octupole moments. These results open a pathway to controlling and transporting electric and magnetic dipole and multipole properties.

\section{Multiferrons}
We begin by calculating the finite electric polarization and magnetization that give rise to multiferrons. LiNbO$_3$ is a ferroelectric semiconductor crystallizing in the rhombohedral $R3c$ space group~\cite{abrahamsFerroelectricLithiumNiobate1966} that exhibits an optical band gap of \SI{3.7}{\eV}~\cite{dharOpticalPropertiesReduced1990}. Its primitive unit cell consists of ten atoms, resulting in 30 phonon modes characterized by the irreducible representations $A_1$, $A_2$, and $E$. We determine the structural and electronic properties of LiNbO$_3$ using density functional theory, the details of which can be found in Note 1 in Supplemental Material (SM)~\cite{supplementalMaterial} (see also references~\cite{kresseInitioMolecularDynamics1993, kresseInitioMoleculardynamicsSimulation1994, kresseEfficientIterativeSchemes1996, kresseEfficiencyAbinitioTotal1996, kresseUltrasoftPseudopotentialsProjector1999, perdewRestoringDensityGradientExpansion2008, monkhorstSpecialPointsBrillouinzone1976, togoFirstprinciplesPhononCalculations2023, togoImplementationStrategiesPhonopy2023, ridahTemperatureDependenceRaman1997, chowdhuryNeutronInelasticScattering1978, kokanyanTemperatureDependenceRaman2014, king-smithTheoryPolarizationCrystalline1993, wempleRelationshipLinearQuadratic1968, glassLowtemperatureBehaviorSpontaneous1976, zhangComparativeFirstprinciplesStudies2017, ceperleyGroundStateElectron1980, perdewGeneralizedGradientApproximation1996, liuGenerationNarrowbandHighintensity2017, liuPumpFrequencyResonances2020, lindemannUberBerechnungMolekularer1910, sokolowski-tintenFemtosecondXrayMeasurement2003} therein).

\subsection{Electric polarization}

We model the ferroelectric polarization $P$ of LiNbO$_3$ by a double-well potential (Fig.~\ref{fig-1:overview}a) in the Landau-Devonshire theory as
\begin{equation}
    \label{eq:double_well}
        V(P) = \alpha P^{2} + \beta P^{4} + \gamma P^{6},
    \end{equation}
with $\alpha < 0$ and $\beta,\gamma > 0$, yielding minima at the equilibrium ferroelectric polarization of $P_{0} = \pm \SI{79.9}{\micro\coulomb\per\cm\squared}$. Owing to the intrinsic anharmonicity at these minima, excitations along the polarization coordinate carry a net polarization and are therefore referred to as ferrons~\cite{tangExcitationsFerroelectricOrder2022}. These excitations can be decomposed into the infrared-active (IR-active) $A_1$ modes, with the biggest contribution coming from the low-frequency mode at \SI{7.16}{\THz}~\cite{mankowskyUltrafastReversalFerroelectric2017, henstridgeNonlocalNonlinearPhononics2022} (SM Note 1). We therefore express the polarization in terms of the phonon amplitude $Q$ as
\begin{equation}
    \label{eq:mode_polarization}
    \mathbf{P}_\mathrm{ph} = \mathbf{Z} Q / V_c, 
\end{equation}
where $V_c$ is the unit cell volume, and $\mathbf{Z} = \sum_{n} \mathbf{Z}^{*}_{n} \frac{\mathbf{q}_{n}}{\sqrt{\mathcal{M}_{n}}}$ is the mode effective charge given by the Born effective charge tensor $\mathbf{Z}^{*}_{n}$, the phonon eigenvector $\mathbf{q}_{n}$, and the mass $\mathcal{M}_{n}$ of atom $n$. Substituting Eq.~\eqref{eq:mode_polarization} into Eq.~\eqref{eq:double_well} yields an equivalent double-well potential in terms of $Q$, which provides a natural starting point for analyzing lattice excitations in LiNbO$_3$. In the following, we will derive the net electric polarization and magnetization generated by multiferrons. 

Close to the minima at $\pm P_0$, the potential energy of a mode can be described using a reduced potential energy expression with the form
\begin{equation}
    \label{eq:anharmonic_pes}
    V (Q) = \frac{\omega^{2}}{2} Q^{2} + a Q^{3}
\end{equation}
with phonon amplitude $Q$, angular frequency $\omega$, and anharmonicity $a$. This one-dimensional potential applies to the $A_1$ modes, as well as to the $E$ modes along the high-symmetry directions (SM Note 2). The cubic term acts as a self-rectification of the mode, for which we obtain a net phonon displacement  $\overline{Q} = - \frac{3 a}{\omega^{2}} \langle Q^{2} \rangle$, leading to a net polarization given by
\begin{equation}
    \label{eq:mode_displacement}
    \overline{\mathbf{P}} = - \frac{\mathbf{Z}}{V_{c}} \frac{3 a}{\omega^{2}} \langle Q^{2} \rangle,
\end{equation}
where $\langle Q^{2} \rangle = \hbar/\omega$ is the mean-squared amplitude of a single phonon. The polarization associated with a single phonon per unit cell, obtained for linear excitations of the $A_1$ and $E$ modes using Eq.~\eqref{eq:mode_displacement}, is summarized in SM Note 3. Excitations of the $A_1$ modes decrease the magnitude of the total polarization (Fig.~\ref{fig-1:overview}a), whereas excitations of the $E$ modes lead to an increase and tilting of the total polarization (Fig.~\ref{fig-1:overview}b), which could be referred to as Higgs-like and Goldstone-like ferrons, respectively~\cite{bauerPolarizationTransportFerroelectrics2023}.

\subsection{Magnetization}

When the degenerate $E$ modes are excited circularly, the resulting ionic motion follows circular or elliptical trajectories. Such motion results in a temporally rotating electric polarization $\mathbf{P}_\mathrm{ph} $, which gives rise to an out-of-plane net magnetization $\overline{\mathbf{M}}$ via the mechanism of dynamical multiferroicity~\cite{juraschekDynamicalMultiferroicity2017, juraschekOrbitalMagneticMoments2019} (Fig.~\ref{fig-1:overview}c), as given by
\begin{equation}
    \label{eq:dynamic_multiferroic_effect}
    \overline{\mathbf{M}} \propto \mathbf{P}_\mathrm{ph}  \times \partial_{t} \mathbf{P}_\mathrm{ph} .
\end{equation}
Microscopically, the effect originates from atomistic electromagnetic loops created by the circular motion of the ions, which give rise to phonon modes carrying a net magnetization, even in nonmagnetic materials~\cite{juraschekDynamicalMultiferroicity2017, juraschekOrbitalMagneticMoments2019, chengLargeEffectivePhonon2020, geilhufeDynamicallyInducedMagnetism2021, xiaoAdiabaticallyInducedOrbital2021, renPhononMagneticMoment2021, baydinMagneticControlSoft2022, hernandezObservationInterplayPhonon2023, geilhufeElectronMagneticMoment2023, zhangGateTunablePhononMagnetic2023, shabalaPhononInverseFaraday2024, kleblUltrafastPseudomagneticFields2025, mustafaOriginLargeEffective2025, chenGaugeTheoryGiant2025, paivaDynamicallyInducedMultiferroic2025, wangInitioTheoryPhonon2025}. In experiment, the mechanism has been confirmed to generate giant effective magnetic fields corresponding to this magnetization~\cite{luoLargeEffectiveMagnetic2023, basiniTerahertzElectricfielddrivenDynamical2024, daviesPhononicSwitchingMagnetization2024, biggsUltrafastFaradayRotation2025}. For a pair of degenerate modes, $\alpha$ and $\beta$, the magnetization associated with a single circularly polarized phonon per unit cell can be written as $\overline{\mathbf{M}} = \frac{\hbar}{V_{c}} \sum_{n} \frac{e \mathbf{Z}^{*}_{n}}{2 \mathcal{M}_{n}} \left( \mathbf{q}_{\alpha,n} \times \mathbf{q}_{\beta,n} \right)$, where $e$ is the elementary charge, and $\mathbf{q}_{\alpha, n}$ and $\mathbf{q}_{\beta, n}$ are the normalized mode eigenvectors of atom $n$ in modes $\alpha$ and $\beta$. The resulting magnetization values $|\overline{\mathbf{M}}|$ for a single circularly polarized phonon are provided in SM Note 3.

\begin{figure}[t]
    \includegraphics{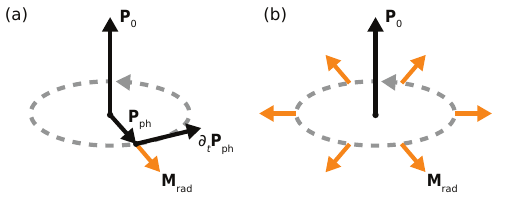}
    \caption{Magnetization induced by circular and elliptical excitation of the $E$ modes in LiNbO$_3$. (a) Superposition between the static ferroelectric polarization $\mathbf{P}_{0}$ and rotating phonon polarization $\partial_{t} \mathbf{P}_\mathrm{ph} $. (b) Resulting radial magnetization $\mathbf{M}_{\mathrm{rad}} $.}
    \label{fig-2:radial_magnetization}
\end{figure}

In addition to the out-of-plane magnetization arising from the circular or elliptical excitation of degenerate modes, the rotating polarization $\mathbf{P}_\mathrm{ph} $ also contains a component perpendicular to the ferroelectric polarization of the material $\mathbf{P}_{0}$, which produces a radial magnetization component given by
\begin{equation}
    \label{eq:dynamic_multiferroic_effect_ferroelectric}
    \mathbf{M}_{\mathrm{rad}}  \propto \mathbf{P}_{0} \times \partial_{t} \mathbf{P}_\mathrm{ph} .
\end{equation}
$\mathbf{M}_{\mathrm{rad}} $ is perpendicular to both $\mathbf{P}_{0}$ and the time derivative of $\mathbf{P}_\mathrm{ph} $, as shown in Fig.~\ref{fig-2:radial_magnetization}. Because the out-of-plane magnetization and radial magnetization both scale with the magnitude of the ionic charge current, the contribution of each atom to the radial magnetization can be expressed as
\begin{equation}
    \label{eq:magnetization_radial_atomic}
    | \mathbf{M}_{\mathrm{rad},n}  | = | \overline{\mathbf{M}}_{n}  | \frac{ | \mathbf{P}_{0} | }{ | \mathbf{P}_{\mathrm{ph},n}  | }
\end{equation}
where $| \mathbf{P}_{\mathrm{ph},n}  |$ and $| \overline{\mathbf{M}}_{n} |$ denote the atomic contributions to the phonon-induced polarization and magnetization, respectively. The radial magnetization is then obtained by summing over all atomic contributions, $\mathbf{M}_{\mathrm{rad}} = \sum_{n} | \mathbf{M}_{\mathrm{rad},n} | \hat{\mathbf{M}}_{\mathrm{rad},n} $, where $\hat{\mathbf{M}}_{\mathrm{rad},n} $ is the unit vector indicating the direction of each atomic contribution.

Interestingly, elliptical excitations of $E$ modes produce both a net polarization and magnetization (Fig.~\ref{fig-1:overview}c). Their elliptical trajectories generate a net out-of-plane magnetization $\overline{\mathbf{M}}$, while simultaneously probing the anharmonicity of the potential landscape, which produces a net in-plane polarization $\overline{\mathbf{P}}$. We accordingly term these excitations \textit{multiferrons}.

\begin{figure*}[t]
    \includegraphics{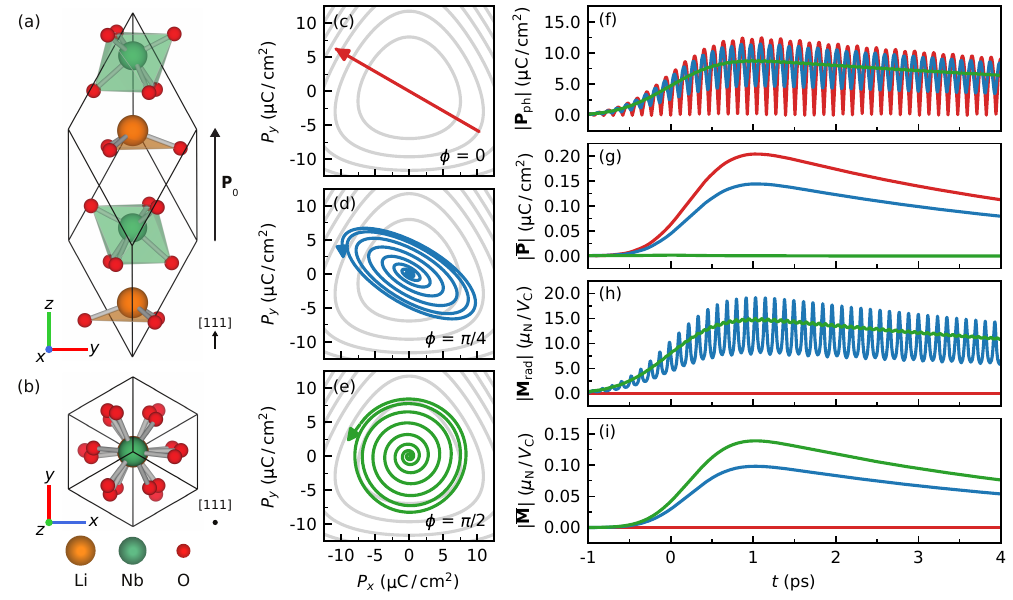}
    \caption{Electric polarization and magnetization dynamics induced by resonant excitation of degenerate $E$ modes at \SI{4.32}{\THz} in LiNbO$_3$. (a,b) Side and top views of the unit cell of LiNbO$_3$. (c-e) Polarization dynamics driven by linearly ($\phi = 0$), elliptically ($\phi = \pi / 4$), and circularly ($\phi = \pi / 2$) polarized pulses in the time window from $-1.0$ to $0.5$~\SI{}{\ps}. Equipotential lines representing the symmetry of the polarization are shown in gray. (f,g) Time evolution of the radial polarization $| \mathbf{P}_{\mathrm{ph}} |$ and net polarization $| \overline{\mathbf{P}} |$. (h,i) Corresponding time evolution of the radial magnetization $|\mathbf{M}_{\mathrm{rad}}|$ and net magnetization $|\overline{\mathbf{M}}|$.}
    \label{fig-3:mode_dynamics}
\end{figure*}


\section{Coherent multiferrons}
%
To study the polarization and magnetization dynamics of multiferrons, we numerically solve the equations of motion for the modes $\alpha \in \{ a, b \}$
\begin{equation}
    \label{eq:phonon_dynamics}
    \begin{split}
        \ddot{Q}_{\alpha} + \kappa_{\alpha} \dot{Q}_{\alpha} + \partial_{Q_{\alpha}} V = \mathbf{Z}_{\alpha} \cdot \mathbf{E} ,
    \end{split}
\end{equation}
where $\kappa_{\alpha} = \omega_{\alpha} / (20 \cdot 2 \pi)$ is the experimental phonon linewidth~\cite{kokanyanTemperatureDependenceRaman2014}, and $V$ the anharmonic potential energy. For two degenerate $E$ modes, the potential energy can be written as
\begin{equation}
    \label{eq:degenerate_pes}
    \begin{split}
        V (Q_{a}, Q_{b}) &= \frac{\omega_{a}^{2}}{2} Q_{a}^{2} + \frac{\omega_{b}^{2}}{2} Q_{b}^{2} + a_{1} Q_{a}^{3} \\
        &+ a_{2} Q_{a}^{2} Q_{b} + a_{3} Q_{a} Q_{b}^{2} + a_{4} Q_{b}^{3},
    \end{split}
\end{equation}
where $\omega_{a} = \omega_{b} \equiv \omega$ and $a_{i}$ $(i = 1, \ldots, 4)$ are the anharmonic coupling coefficients. The right-hand side of Eq.~\eqref{eq:phonon_dynamics} describes the external driving force from a laser pulse, where $\mathbf{Z}_{\alpha}$ is the mode effective charge that determines the coupling to the electric field $\mathbf{E}$. The temporal profile and polarization of the driving field govern the amplitude and phase of the coherent phonon response. The functional form used to model the laser pulse is given in SM Note 4.

In Fig.~\ref{fig-3:mode_dynamics}, we present the dynamics of the degenerate $E$ modes at \SI{4.32}{\THz} in LiNbO$_3$, with other $E$ modes shown in SM Note 5. Figures~\ref{fig-3:mode_dynamics}a,b show the unit cell from two different crystallographic directions, where the $E$ modes are polarized in the $xy$ plane. To demonstrate the multiferron dynamics, we excite the low-frequency modes using linear, elliptical, and circular laser pulses, along the high-symmetry lines of the potential energy surface, as shown in Figs.~\ref{fig-3:mode_dynamics}c-e. In all cases, the polarization dynamics follow the polarization of the laser pulse: a linearly polarized pulse leads to linear oscillations of the polarization, whereas elliptically and circularly polarized pulses generate a precessing polarization. This behavior is also reflected by the time evolution of the magnitude of radial polarization $| \mathbf{P}_{\mathrm{ph}} |$ in Fig.~\ref{fig-3:mode_dynamics}f.

We next investigate how the phonon anharmonicity leads to a net polarization $|\overline{\mathbf{P}}|$ upon driving the $E$ modes, which we show in Fig.~\ref{fig-3:mode_dynamics}g. For both linear and elliptical excitations, a nonzero in-plane polarization is generated that tilts the overall polarization of the material. The envelope of the net polarization follows the time-dependent phonon population number, $|\overline{\mathbf{P}}|\propto \langle Q^2\rangle \propto N$, which decays on a timescale determined by the phonon linewidth $\kappa_\alpha$. $|\overline{\mathbf{P}}|$ is largest for linearly polarized driving and reduced by a factor of $\sqrt{2}$ for elliptically polarized driving. Circular excitations probe the anharmonic potential energy surface uniformly and do not create net polarization.

Upon investigating magnetization dynamics, we observe a substantial radial magnetization $| \mathbf{M}_{\mathrm{rad}} |$ for elliptical and circular excitation, which we show in Fig.~\ref{fig-3:mode_dynamics}h. This effect originates from the superposition of the rotating polarization and the ferroelectric polarization, as described by Eq.~\eqref{eq:dynamic_multiferroic_effect_ferroelectric}. Similarly, the net out-of-plane magnetization $| \overline{\mathbf{M}} |$ is only produced by elliptical and circular excitations (Fig.~\ref{fig-3:mode_dynamics}i). The envelope of $|\overline{\mathbf{M}}|$ follows the time-dependent phonon population in the same way as $|\overline{\mathbf{P}}|$. Notably, the magnitude of the radial magnetization is two orders of magnitude larger than that of the net out-of-plane magnetization.

Elliptically polarized driving yields both a net in-plane polarization and a net out-of-plane magnetization as a result of the coherent excitation of multiferrons. The relative strength of the induced electric polarization and magnetization can be tuned continuously by tuning the ellipticity of the laser pulse, as shown in SM Note 5. The multiferrons further generate a rotating radial magnetization and polarization parallel to each other (Fig.~\ref{fig-2:radial_magnetization}). We will show in the following that these rotating contributions additionally lead to the emergence of electric and magnetic multipole moments.

\section{Multipolons}
We next investigate the multipolar nature of the radial polarization and magnetization induced by the multiferrons, which we express using quadrupole tensors $\mathcal{Q}_{ij}$. Within each unit cell, the contributions of the atoms to the quadrupole tensors are given by
\begin{align}
    \label{eq:quadrupole_tensors}
    \langle \mathcal{Q}^{P}_{n,ij} \rangle &= \langle u_{n,i} P_{\mathrm{ph},n,j} \rangle - \frac{1}{3} \delta_{ij} \sum_{k} \langle u_{n,k} P_{\mathrm{ph},n,k} \rangle \\
    \langle \mathcal{Q}^{M}_{n,ij} \rangle &= \langle u_{n,i} M_{\mathrm{rad},n,j} \rangle - \frac{1}{3} \delta_{ij} \sum_{k} \langle u_{n,k} M_{\mathrm{rad},n,k} \rangle
\end{align}
where $\langle \cdots \rangle$ denotes an average over the phonon period, $i,j,k \in \{ x, y, z \}$ indicate the Cartesian components of the vectors, $u_{n,i}  = \sum_{\alpha} Q_{\alpha}  \frac{q_{\alpha,n,i}}{\sqrt{\mathcal{M}_{n}}}$ are the atomic displacements, and $\delta_{ij}$ is the Kronecker delta function. Each total quadrupole tensor component is obtained by summing over all atomic contributions: $\langle \mathcal{Q}_{ij} \rangle = \sum_{n} \langle \mathcal{Q}_{n,ij} \rangle$. In a similar manner, the octupole tensor can be computed, a description of which is provided in SM Note 6. These multipolar tensors are constructed as sums over the individual contributions from the local atomic displacements.

\begin{figure}[t]
    \includegraphics{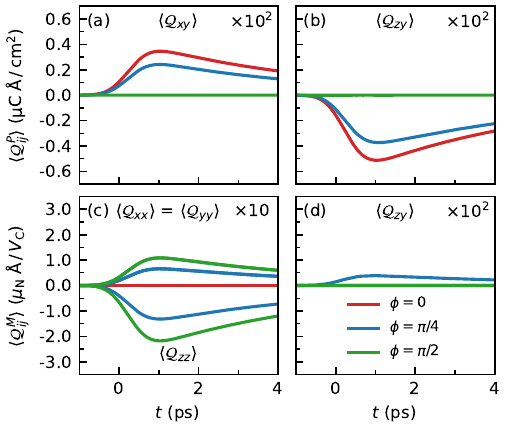}
    \caption{Time-averaged quadrupole tensors of radial polarization (top) and magnetization (bottom) for linearly ($\phi = 0$), elliptically ($\phi = \pi / 4$), and circularly ($\phi = \pi / 2$) polarized pulses. (a,b) Off-diagonal quadrupole tensor contributions to radial polarization $\mathbf{P}_{\mathrm{ph}}$. (c,d) Diagonal and off-diagonal quadrupole tensor contributions to radial magnetization $\mathbf{M}_{\mathrm{rad}}$.}
    \label{fig-4:quadrupole_tensor}
\end{figure}

Figure~\ref{fig-4:quadrupole_tensor} shows the temporal evolution of cycle-averaged components of the polarization and magnetic quadrupole tensors upon excitation by linearly, elliptically, and circularly polarized laser pulses. A complete overview of all components is provided in SM Note 6. Notably, the nonzero $\langle \mathcal{Q}_{xy} \rangle$ component in Fig.~\ref{fig-4:quadrupole_tensor}a can be used to describe the breaking of the axial symmetry of the radial polarization $\mathbf{P}_{\mathrm{ph}}$ for both linear and elliptical excitations. Additionally, the finite value of $\langle \mathcal{Q}_{zy} \rangle \neq 0$ (Fig.~\ref{fig-4:quadrupole_tensor}b) indicates that such excitations also lead to a small out-of-plane tilt of the polarization.

Focusing on the multipole moments of the radial magnetization $\mathbf{M}_{\mathrm{rad}}$, we observe that circular and elliptical excitations also yield substantial nonzero diagonal components to the quadrupole tensor, i.e. $\langle \mathcal{Q}^{M}_{ij} \rangle = 0$ (Fig.~\ref{fig-4:quadrupole_tensor}c). Linear excitations do not yield any diagonal components since they do not give rise to any radial magnetization. Notably, elliptical excitations also give rise to a nonzero $zy$-component in the quadrupole tensor as  seen in Fig.~\ref{fig-4:quadrupole_tensor}d). This off-diagonal term tilts the magnetization distribution out of the transverse plane, breaking the in-plane symmetry that is present for circular excitations.

Our results show that multiferrons can generate structured polarization and magnetization patterns with multipole moments. The induced finite quadrupole and octupole tensor components can be tuned by changing the ellipticity of the incident laser pulse, similar to the net polarization and magnetization components. We therefore term these excitations \textit{multipolons}.


\section{Discussion}
We have performed our calculations for the example of LiNbO$_3$, however, multiferrons and multipolons can be created in all ferroelectric materials hosting doubly degenerate optical phonons. The size of the magnetic moment associated with multiferrons depends on the effective charges of the ions, making oxide perovskites~\cite{juraschekOrbitalMagneticMoments2019, basiniTerahertzElectricfielddrivenDynamical2024} or other materials with large phonon magnetic moments such as Pb$_{1-x}$Sn$_x$Te~\cite{baydinMagneticControlSoft2022, hernandezObservationInterplayPhonon2023} attractive candidates for large multiferronic effects. As the multipolar moments are proportional to the ferroelectric polarization, oxide perovskites are also a promising choice to observe multipolons.

While multiferrons share similarities with magnetoferrons and electromagnons, which are hybrid quasiparticles that appear in multiferroic materials~\cite{castroPhenomenologicalTheoryElectromagnons2025}, they differ in two aspects. First, multiferrons do not require magnetic order to be present. Second, multiferrons carry both a finite polarization and magnetization, whereas the other quasiparticles carry either only one (magnetoferrons) or neither (electromagnons and circularly polarized phonons). In contrast, multiferrons do not require magnetic order to be present. Furthermore, recent predictions show that nonlinear phonon interaction allows for electric polarization and magnetization to be created simultaneously in nonpolar nonmagnetic materials~\cite{paivaDynamicallyInducedMultiferroic2025}. These materials do not host multiferron quasiparticles however, as their symmetry does not allow for $Q^3$-type anharmonicities.

Coherent excitation of multiferrons with an ultrashort laser pulse simultaneously generates macroscopic electric polarization and magnetization and provides a path to controlling multiple ferroic orders at once. In addition, the multipolar character of multiferrons opens possibilities regarding their coupling to external magnetic systems, as magnetic multipolar order plays an important role in orbital magnetism~\cite{shitadeTheoryOrbitalMagnetic2018} and altermagnetism~\cite{bhowalFerroicallyOrderedMagnetic2024}, and has recently been shown to couple to lattice vibrations~\cite{hartPhononDrivenMultipolarDynamics2025, sutcliffePseudochiralPhononSplitting2025}. The magnetic multipoles could be detected directly using inelastic neutron scattering, where signals can be attributed directly to magnetic quadrupoles~\cite{urruNeutronScatteringLocal2023}.

While the coherent excitation of multiferrons with light demonstrated here occurs at the Brillouin-zone center, multiferrons at finite wave vectors can be excited thermally. In ferroelectric materials, IR-active phonon branches carry nonzero angular momentum and magnetic moments~\cite{uedaChiralPhononsPolar2025}. Although polar modes are generally considered in the limit $\mathbf{q} = 0$, transport experiments have shown that ferrons have substantial group velocity~\cite{shenObservationFerronTransport2025, choeObservationCoherentFerrons2025}. Therefore, the pyrocaloric effects which arise from ferrons~\cite{tangExcitationsFerroelectricOrder2022} can be expected to have magnetocaloric complements in multiferrons. A thermal magnetization from multiferrons, like the thermally induced polarization from ferrons, would arise from broken rotational symmetry in the anharmonic potential at $\mathbf{q} \neq 0$. As propagating ferrons can transmit signals over micrometer distances~\cite{shenObservationFerronTransport2025, choeObservationCoherentFerrons2025}, they may find potential applications in information technology. Ferron transport can be created by electric fields and temperature gradients~\cite{tangThermoelectricPolarizationTransport2022}, whereas multiferrons could further couple to magnetic fields, opening new possibilities for creating and manipulating their transport. To understand the transport properties of multiferrons, further study of their behavior at finite wave vectors is therefore required.

During the preparation of this manuscript, we became aware of a closely related study on multiferrons that was recently posted online~\cite{tangMultiferroiclikeQuasiparticlesFerroelectrics2025}.

\section*{Data availability}

The data that support the findings of this article are openly available in Ref.~\cite{zenodoData2026}.

\begin{acknowledgments}

We thank Michael Fechner for useful discussions. This research was supported by the ERC Starting Grant CHIRALPHONONICS, No. 101166037 C.P.R. acknowledges support from the project {FerrMion} of the Ministry of Education, Youth and Sports, Czech Republic, co-funded by the European Union (CZ.02.01.01/00/22\_008/0004591).

\end{acknowledgments}

\bibliography{ref}

\end{document}


\title{Supplemental Material: \\ Lattice Excitations with Finite Polarization and Magnetization}

\author{Mike Pols}
\email{mike.pols@mat.ethz.ch}
\altaffiliation{%
    Present address: Materials Theory, ETH Zurich, CH-8093 Z\"urich, Switzerland
}%
\affiliation{%
    Department of Applied Physics and Science Education, Eindhoven University of Technology, 5612 AP Eindhoven, Netherlands
}%

\author{Carl P. Romao}
\email{carl.romao@cvut.cz}
\affiliation{%
    Department of Materials, Faculty of Nuclear Sciences and Physical Engineering, Czech Technical University in Prague, Trojanova 13, Prague 120 00, Czech Republic
}%

\author{Dominik M. Juraschek}
\email{d.m.juraschek@tue.nl}
\affiliation{%
    Department of Applied Physics and Science Education, Eindhoven University of Technology, 5612 AP Eindhoven, Netherlands
}%

\date{\today}

\maketitle

\clearpage

\tableofcontents

\clearpage

\section{Computational details}

\subsection*{Density functional theory (DFT)}

We calculate the properties of LiNbO$_{3}$ using density functional theory (DFT) as implemented in the Vienna \emph{Ab-initio} Simulation Package (\texttt{VASP})~\cite{kresseInitioMolecularDynamics1993, kresseInitioMoleculardynamicsSimulation1994, kresseEfficientIterativeSchemes1996, kresseEfficiencyAbinitioTotal1996}. We use the standard PAW pseudopotentials with the valence electron configurations Li (2s$^{1}$), Nb (4p$^{6}$5s$^{1}$4d$^{4}$), and O (2s$^{2}$2p$^{4}$)~\cite{kresseUltrasoftPseudopotentialsProjector1999}. Exchange-correlation interactions are described using the PBEsol functional~\cite{perdewRestoringDensityGradientExpansion2008}. Convergence is achieved with an energy cutoff of \SI{600}{\eV} and a \kpoints{6}{6}{6} \textGamma-centered $\mathbf{k}$-mesh~\cite{monkhorstSpecialPointsBrillouinzone1976}. The geometry is optimized using energy and force convergence criteria of \SI{1E-9}{\eV} and \SI{1E-5}{\eV\per\angstrom}, respectively. Our optimized hexagonal cell, with lattice parameters $a = \SI{5.16}{\angstrom}$ and $c = \SI{13.93}{\angstrom}$, fits well to the experimental cell in Ref.~\cite{abrahamsFerroelectricLithiumNiobate1966} (Table~\ref{tab:xc_functionals}).

\subsection*{Phonon calculations}

We compute the second-order interatomic force constants (IFCs) of LiNbO$_{3}$ in a \supercell{3}{3}{3} supercell using the real-space supercell method as implemented in \texttt{Phonopy}~\cite{togoFirstprinciplesPhononCalculations2023, togoImplementationStrategiesPhonopy2023}. These IFCs are obtained by displacing the atoms \SI{0.01}{\angstrom} from their equilibrium positions. The resulting real-space IFCs form the basis for computing phonon frequencies and eigenvectors.

To assess the accuracy of the computed phonons, we compare the resulting frequencies against those obtained from density functional perturbation theory (DFPT) and available experimental data. The comparison, shown in Table~\ref{tab:phonon_frequencies}, demonstrates excellent agreement among the different methods. In particular, the DFT-based finite displacement method reproduces DFPT frequencies within \SI{0.05}{\THz}, and both match well with experimental results for $A_1$ modes~\cite{ridahTemperatureDependenceRaman1997}, $A_2$ modes~\cite{chowdhuryNeutronInelasticScattering1978}, and $E$ modes~\cite{kokanyanTemperatureDependenceRaman2014}. This agreement indicates that the computed IFCs and derived phonons accurately describe the lattice dynamics of LiNbO$_{3}$.

\begin{table*}[htbp!]
\caption{\label{tab:phonon_frequencies}Frequencies $\nu_{0}$ of optical phonon modes in LiNbO$_{3}$, calculated with density functional perturbation theory (DFPT) and the finite displacement method with density functional theory (DFT). Experimental frequencies from Refs.~\cite{ridahTemperatureDependenceRaman1997, chowdhuryNeutronInelasticScattering1978, kokanyanTemperatureDependenceRaman2014}}
\begin{ruledtabular}
    \begin{tabular}{llccc}
    Mode & Symm. & $\nu^{\mathrm{DFPT}}_{0}$ (\SI{}{\THz}) & $\nu^{\mathrm{DFT}}_{0}$ (\SI{}{\THz}) & $\nu^{\mathrm{Exp.}}_{0}$ (\SI{}{\THz}) \\
    \colrule
    1, 2, 3 & -             & 0.00      & 0.00      & -         \\
    4, 5    & $E$           & 4.32      & 4.32      & 4.54      \\
    6       & $A_2$         & 6.25      & 6.26      & 6.70      \\
    7, 8    & $E$           & 6.42      & 6.40      & -         \\
    9       & $A_1$         & 7.17      & 7.16      & 7.55      \\
    10, 11  & $E$           & 7.61      & 7.61      & 7.08      \\
    12      & $A_1$         & 8.02      & 8.05      & 8.24      \\
    13      & $A_2$         & 8.47      & 8.48      & 9.42      \\
    14, 15  & $E$           & 9.43      & 9.44      & -         \\
    16      & $A_1$         & 10.07     & 10.08     & 9.95      \\
    17, 18  & $E$           & 10.50     & 10.54     & 9.63      \\
    19, 20  & $E$           & 10.85     & 10.88     & 11.05     \\
    21      & $A_2$         & 11.87     & 11.85     & -         \\
    22, 23  & $E$           & 12.34     & 12.35     & -         \\
    24      & $A_2$         & 13.11     & 13.11     & 13.65     \\
    25, 26  & $E$           & 17.04     & 17.04     & 17.34     \\
    27      & $A_1$         & 18.24     & 18.24     & 18.95     \\
    28, 29  & $E$           & 19.93     & 19.92     & -         \\
    30      & $A_2$         & 26.10     & 26.11     & -         \\
    \end{tabular}
\end{ruledtabular}
\end{table*}

\clearpage

\subsection*{Ferroelectric polarization}

\begin{figure*}[htbp!]
    \includegraphics{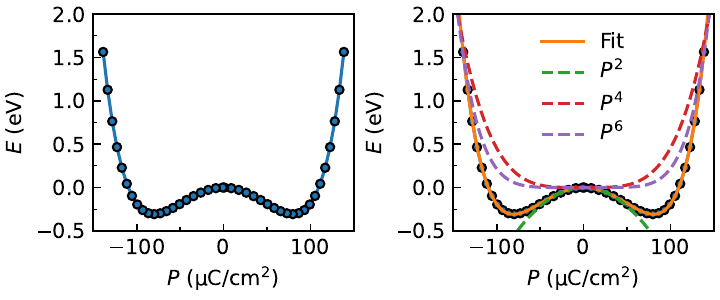}
    \caption{(a) Double-well potential energy landscape for ferroelectric polarization in LiNbO$_{3}$. (b) Fitted double-well potential with the total potential (orange) and individual contributions (green, red, and purple) plotted separately.}
    \label{fig-si:polarization-landscape}
\end{figure*}

To calculate the polarization of the material, we generate a series of 21 structures that interpolates between ferroelectric LiNbO$_{3}$ ($R3c$; \#161) and its inverse structure, passing through a paraelectric structure ($R\overline{3}c$; \#167)~\cite{abrahamsFerroelectricLithiumNiobate1966}. At both ends we extrapolate beyond the ferroelectric structures with an additional 10 structures of similar step size. In total, this yields 41 structures with which we capture the double-well character of the potential. For each structure, we calculate the polarization according to the modern theory of polarization~\cite{king-smithTheoryPolarizationCrystalline1993}. The calculated polarization of LiNbO$_{3}$ is $\mathbf{P}_{0} = \SI{79.9}{\micro\coulomb\per\cm\squared}$, matching experimental~\cite{wempleRelationshipLinearQuadratic1968, glassLowtemperatureBehaviorSpontaneous1976} and computational results~\cite{zhangComparativeFirstprinciplesStudies2017} from literature. The full polarization landscape is shown in Fig.~\ref{fig-si:polarization-landscape}.

For insights into the contributions of the various modes to the transition between the two ferroelectric states along the double-well potential, we project the structural distortions onto the $A_1$ modes. In doing so, the displacement of each atom, $\Delta \mathbf{r}_{n}$, can be expressed as a sum over mode contributions as
\begin{equation}
    \label{eq:distortion-projection}
    \Delta \mathbf{r}_{n} = \mathbf{r}_{\mathrm{PE},n} - \mathbf{r}_{\mathrm{FE},n} = \sum_{\alpha} c_{\alpha} \frac{\mathbf{q}_{\alpha, n}}{\sqrt{\mathcal{M}_{n}}}
\end{equation}
where $\mathbf{r}_{\mathrm{PE},n}$ and $\mathbf{r}_{\mathrm{FE},n}$ are the atomic coordinates in the paraelectric and ferroelectric phases of LiNbO$_{3}$, respectively. The coefficients $c_{\alpha}$ quantify the contribution of each mode, $\mathbf{q}_{\alpha, n}$ is the normalized eigenvector of atom $n$ in mode $\alpha$, and $\mathcal{M}_{n}$ is the atomic mass. 

An overview of the mode contributions can be found in Table~\ref{tab:mode_contributions}. The lowest-energy $A_1$ mode dominates both the structural distortion and, due to its large mode effective charge, the net polarization change, while the other $A_1$ modes only provide minor contributions. This establishes the mode at \SI{7.16}{\THz} as the primary mode for ferrons in LiNbO$_{3}$.

\begin{table*}[htbp!]
\caption{\label{tab:mode_contributions}Contribution coefficients of the $A_1$ modes to the ferroelectric to paraelectric transition $c_{\alpha}$, along with their mode effective charges $\mathbf{Z}$ and fractional contributions to the polarization change, i.e. $| c_{\alpha} \, \mathbf{Z} | / \sum_{\alpha} | c_{\alpha} \, \mathbf{Z} | \cdot 100\%$.}
\begin{ruledtabular}
    \begin{tabular}{ccccc}
    Mode & $\nu^{\mathrm{DFT}}_{0}$ (\SI{}{\THz}) & $| c_{\alpha} |$ (-) & $| \mathbf{Z} |$ (\SI{}{\elementarycharge\per{\sqrt{\amu}}}) & $| c_{\alpha} \, \mathbf{Z} | / \sum_{\alpha} | c_{\alpha} \, \mathbf{Z} | \cdot 100\%$ \\
    \colrule
    9       & 7.16      & 2.73      & 1.512     & 86.1\%    \\
    12      & 8.05      & 1.07      & 0.118     &  2.6\%    \\
    16      & 10.08     & 0.06      & 0.287     &  0.4\%    \\
    27      & 18.24     & 0.32      & 1.629     & 10.9\%    \\
    \end{tabular}
\end{ruledtabular}
\end{table*}

\subsection*{Exchange-correlation functionals}

To evaluate the effect that different exchange-correlation functionals have on the observables of LiNbO$_3$, we also optimize its geometry and compute the ferroelectric polarization using the LDA~\cite{ceperleyGroundStateElectron1980} and PBE~\cite{perdewGeneralizedGradientApproximation1996} functionals. The overview of these calculations, along with experimental reference values, is shown in Table~\ref{tab:xc_functionals}. All functionals reproduce the geometry and ferroelectric polarization well.

\begin{table*}[htbp!]
\caption{\label{tab:xc_functionals}Optimized lattice parameters $a$ and $c$ as well as the ferroelectric polarization $| \mathbf{P}_{0} |$ for LDA, PBE, and PBEsol exchange-correlation functional. Experimental values are added for reference.}
\begin{ruledtabular}
    \begin{tabular}{cccc}
    & $a$ (\SI{}{\angstrom}) & $c$ (\SI{}{\angstrom}) & $| \mathbf{P_{0}} |$ (\SI{}{\micro\coulomb\per\cm\squared}) \\
    \colrule
    Exp.    & 5.15~\cite{abrahamsFerroelectricLithiumNiobate1966} & 13.86~\cite{abrahamsFerroelectricLithiumNiobate1966} & 70-71~\cite{wempleRelationshipLinearQuadratic1968, glassLowtemperatureBehaviorSpontaneous1976} \\
    LDA     & 5.10      & 13.81     & 77.1  \\
    PBE     & 5.21      & 14.12     & 84.0  \\
    PBEsol  & 5.16      & 13.93     & 79.9  \\
    \end{tabular}
\end{ruledtabular}
\end{table*}

\clearpage

\section{Phonon potential energy surfaces (PES)}

\subsection*{Transforming degenerate PES to cylindrical coordinates}

We model the potential energy surface (PES) of two degenerate $E$ modes in LiNbO$_{3}$ as
\begin{equation}
    \label{eq:total_pes}
    V ( Q_{a}, Q_{b} ) = \frac{\omega_{a}^2}{2} Q_{a}^{2} + \frac{\omega_{b}^2}{2} Q_{b}^{2} + a_{1} Q_{a}^{3} + a_{2} Q_{a}^{2} Q_{b} + a_{3} Q_{a} Q_{b}^{2} + a_{4} Q_{b}^{3}
\end{equation}
where $Q_{a}$ and $Q_{b}$ are the mode amplitudes of the degenerate modes, $\omega_{a} = \omega_{b} \equiv \omega$ is the phonon frequency, and $a_{i}$ $(i = 1, \ldots, 4)$ are the anharmonic coupling coefficients. The PES can be separated into a harmonic and anharmonic contribution as
\begin{align}
    \label{eq:2nd_pes}
    V^{(2)} ( Q_{a}, Q_{b} ) &= \frac{\omega^2}{2} Q_{a}^{2} + \frac{\omega^2}{2} Q_{b}^{2} \\
    \label{eq:3rd_pes}
    V^{(3)} ( Q_{a}, Q_{b} ) &= a_{1} Q_{a}^{3} + a_{2} Q_{a}^{2} Q_{b} + a_{3} Q_{a} Q_{b}^{2} + a_{4} Q_{b}^{3}
\end{align}
of which $V^{(2)}$ has complete rotational symmetry and $V^{(3)}$ a 3-fold rotational symmetry, as a result of the $R3c$ space group of LiNbO$_{3}$~\cite{abrahamsFerroelectricLithiumNiobate1966}.

To explicitly show this 3-fold rotational symmetry in $V^{(3)}$, we rewrite it in cylindrical mode coordinates. We substitute
\begin{equation}
    \label{eq:cylindrical_mode_coordinates}
    Q_{a} = Q \cosn{}{\theta} \; \text{and} \; Q_{b} = Q \sinn{}{\theta}
\end{equation}
into Eq.~\eqref{eq:3rd_pes}, resulting in the following relation for the anharmonic PES
\begin{equation}
    \label{eq:3rd_pes_cylindrical}
    V^{(3)} ( Q, \theta ) = Q^{3} \Big [ a_{1} \cosn{3}{\theta} + a_{2} \cosn{2}{\theta} \sinn{}{\theta} + a_{3} \cosn{}{\theta} \sinn{2}{\theta} + a_{4} \sinn{3}{\theta} \Big ].
\end{equation}
This expression can be further simplified with the following trigonometric relations
\begin{equation}
    \label{eq:trigonometric_relations}
    \begin{split}
        \cosn{3}{\theta}                    &= \frac{3}{4} \cosn{}{\theta} + \frac{1}{4} \sinn{}{3 \theta} \\
        \sinn{3}{\theta}                    &= \frac{3}{4} \sinn{}{\theta} - \frac{1}{4} \cosn{}{3 \theta} \\
        \cosn{2}{\theta} \sinn{}{\theta}    &= \frac{1}{4} \sinn{}{\theta} + \frac{1}{4} \sinn{}{3 \theta} \\
        \sinn{2}{\theta} \cosn{}{\theta}    &= \frac{1}{4} \cosn{}{\theta} - \frac{1}{4} \cosn{}{3 \theta}
    \end{split}
\end{equation}
resulting in the following relation
\begin{equation}
    \label{eq:3rd_pes_cylindrical_1}
    \begin{split}
        V^{(3)} ( Q, \theta ) = Q^{3} \Bigg[ &\left( \frac{3}{4} a_{1} + \frac{1}{4} a_{3} \right) \cosn{}{\theta} + \left( \frac{3}{4} a_{4} + \frac{1}{4} a_{2} \right) \sinn{}{\theta} \\
        + &\left( \frac{1}{4} a_{1} - \frac{1}{4} a_{3} \right) \cosn{}{3 \theta} + \left( - \frac{1}{4} a_{4} + \frac{1}{4} a_{2} \right) \sinn{}{3 \theta} \Bigg].
    \end{split}
\end{equation}

Noting the relationship between the coupling coefficients
\begin{equation}
    \label{eq:coupling_coefficients}
    -3 a_{1} = a_{3} \; \text{and} \; -3 a_{4} = a_{2},
\end{equation}
the expression for the anharmonic PES can be further simplified to
\begin{equation}
    \label{eq:3rd_pes_cylindrical_2}
    V^{(3)} ( Q, \theta ) = Q^{3} \Big[ a_{1} \cosn{}{3 \theta} - a_{4} \sinn{}{3 \theta} \Big]
\end{equation}
for which the 3-fold symmetry is present due to the $\cosn{}{3 \theta}$ and $\sinn{}{3 \theta}$ terms. Ultimately, we can rewrite this expression to
\begin{equation}
    \label{eq:3rd_pes_cylindrical_3}
    \begin{split}
        V^{(3)} ( Q, \theta )   &= a' Q^{3} \Big[ \cosn{}{3 \theta + \delta} \Big] \\
        a'                  &= \sqrt{ \left| a_{1} \right|^{2} + \left| a_{4} \right|^{2}} \\
        \delta              &= \atantwo \left( -\frac{a_{4}}{a_{1}} \right)
    \end{split}
\end{equation}
where $\delta$ is the offset of the 3-fold angular modulation in the PES, which can be used to compute the first minimum in the modulation as
\begin{equation}
    \label{eq:theta_minimum}
    3 \theta_{\min} + \delta = \pi \rightarrow \theta_{\min} = \frac{\pi - \delta}{3}.
\end{equation}
Along this high-symmetry direction, i.e. $\theta = \theta_{\min}$, the potential energy of the degenerate modes reduces to
\begin{equation}
    \label{eq:high_symmetry_pes}
    V ( Q ) = \frac{\omega^{2}}{2} Q^{2} + a' Q^{3},
\end{equation}
showing similarity to the anharmonic potential of the $A_1$ modes.

\begin{figure*}[htbp!]
    \includegraphics{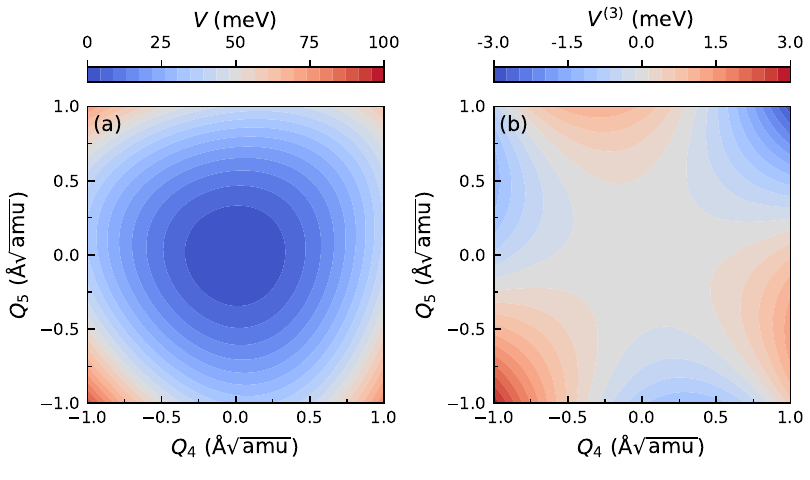}
    \caption{Potential energy landscape of the degenerate $E$ modes at \SI{10.54}{\THz}. (a) Total potential energy surface $V$ with the anharmonic contribution multiplied by a factor of 5. (b) Anharmonic contribution $V^{(3)}$. Both potential energy surfaces exhibit a 3-fold rotational symmetry.}
    \label{fig-si:anharmonic_pes}
\end{figure*}

In Fig.~\ref{fig-si:anharmonic_pes}, we present the total potential energy surface and the anharmonic contribution for the degenerate modes at \SI{10.54}{\THz}. The surface shows a 3-fold symmetry, consistent with the symmetry of LiNbO$_{3}$.

\FloatBarrier

\subsection*{Phonon anharmonicity}

To compute the anharmonicity of the phonons, we use the brute-force (BF) method, which involves displacing atoms along phonon eigenvectors and extracting the coupling coefficients from the resulting changes in total energy. In this work, we apply uniform displacements to the atoms of LiNbO$_{3}$ along each mode. Displacements are sampled on a 9-point grid spanning $\pm$\SI{1.0}{\angstrom \sqrt{amu}}, resulting in 9 structures for the $A_1$ modes and 81 structures for each pair of degenerate $E$ modes. An overview of the anharmonicity of the various modes is provided in Table~\ref{tab:phonon_couplings}.

For the $A_1$ modes, we fit the potential energy of the displaced structures to the form
\begin{equation}
    \label{eq:a_mode_potential}
    V_{A_{1}} ( Q ) = \frac{\omega^{2}}{2} Q^{2} + a Q^{3},
\end{equation}
where $\omega$ is fixed using the harmonic phonon frequency and $a$ quantifies the anharmonicity.

For the degenerate $E$ modes, the potential energy depends on the mode amplitudes $Q_{a}$ and $Q_{b}$ and is fit as
\begin{equation}
    \label{eq:e_mode_potential}
    \begin{split}
        V_{E} ( Q_{a}, Q_{b} ) &= \frac{\omega_{a}^{2}}{2} Q_{a}^{2} + \frac{\omega_{b}^{2}}{2} Q_{b}^{2} + a_{1} Q_{a}^{3} + a_{2} Q_{a}^{2} Q_{b} + a_{3} Q_{a} Q_{b}^{2} + a_{4} Q_{b}^{3} \\
        &= \frac{\omega_{a}^{2}}{2} Q_{a}^{2} + \frac{\omega_{b}^{2}}{2} Q_{b}^{2} + a_{1} \left( Q_{a}^{3} - 3 Q_{a} Q_{b}^{2} \right) + a_{4} \left( Q_{b}^{3} - 3 Q_{a}^{2} Q_{b} \right)
    \end{split}
\end{equation}
where $\omega_{a} = \omega_{b} = \omega$ is the harmonic phonon frequency of the degenerate modes, and $a_{i}$ $(i = 1, \ldots, 4)$ are the anharmonic coupling coefficients. The cubic invariants $Q_{a}^{3} - 3 Q_{a} Q_{b}^{2}$ and $Q_{b}^{3} - 3 Q_{a}^{2} Q_{b}$ enforce the symmetry constraints imposed by the point group operations on the $E$ modes. These invariants follow directly from the relations in Eq.~\eqref{eq:coupling_coefficients}. The overall anharmonic coupling strength is then given by $a'= \sqrt{| a_{1} |^{2} + | a_{4} |^{2}}$ as in Eq.~\eqref{eq:3rd_pes_cylindrical_3}.

\begin{table*}[htbp!]
\caption{\label{tab:phonon_couplings}Magnitude of anharmonic phonon coupling coefficients, $| a |$ or $| a' |$, calculated using the brute-force (BF) method.}
\begin{ruledtabular}
    \begin{tabular}{llcc}
    Symmetry & Mode & $\nu^{\mathrm{DFT}}_{0}$ (\SI{}{\THz}) & $| a |$ or $| a' |$ $\left( \SI{}{\milli\eV\per(\angstrom\sqrt{\text{amu}})\cubed} \right)$ \\
    \colrule
    $A_1$       & 9         &      7.16 &     18.72 \\
                & 12        &      8.05 &     19.02 \\
                & 16        &     10.08 &      6.83 \\
                & 27        &     18.24 &    168.98 \\
    \colrule
    $E$         & 4, 5      &      4.32 &      1.05 \\
                & 7, 8      &      6.40 &      1.35 \\
                & 10, 11    &      7.61 &      5.65 \\
                & 14, 15    &      9.44 &      5.03 \\
                & 17, 18    &     10.54 &     33.08 \\
                & 19, 20    &     10.88 &      7.13 \\
                & 22, 23    &     12.35 &      4.06 \\
                & 25, 26    &     17.04 &     60.23 \\
                & 28, 29    &     19.92 &      1.83 \\
    \end{tabular}
\end{ruledtabular}
\end{table*}

\clearpage

\begin{figure*}[htbp!]
    \includegraphics{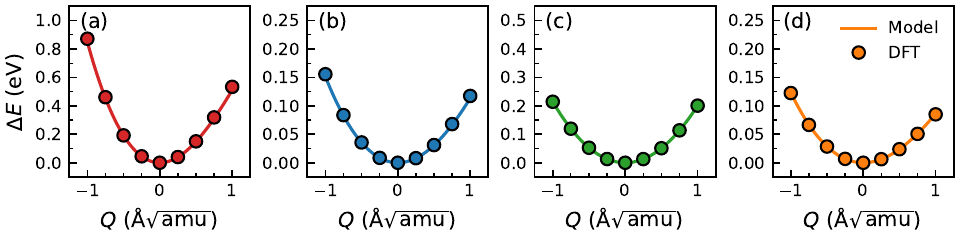}
    \caption{Comparison of the cubic approximation (model) to the potential energy for the $A_1$ modes from density functional theory (DFT) calculations. Potential energies of (a) mode 9 (\SI{7.16}{\THz}), (b) mode 12 (\SI{8.05}{\THz}), (c) mode 16 (\SI{10.08}{\THz}), and (d) mode 27 (\SI{18.24}{\THz}) are shown.}
    \label{fig-si:cubic_approximation}
\end{figure*}

In Fig.~\ref{fig-si:cubic_approximation}, we present an overview of the cubic approximation to the potential energy of the $A_1$ modes. As shown, the cubic approximation from Eq.~\eqref{eq:a_mode_potential} accurately reproduces the potential energy in the range of $-1.0$ to \SI{+1.0}{\angstrom \sqrt{amu}}.

\clearpage

\section{Single-phonon per unit cell properties}

\subsection*{Single-phonon per unit cell calculations}

Treating phonons as a quantum mechanical harmonic oscillator, we can relate the mean-squared displacement of a mode $\langle Q^{2} \rangle$ to the angular frequency $\omega$ of that mode using ladder operators, $a$ and $a^{\dagger}$, as
\begin{equation}
    \label{eq:single_phonon_displacement}
    \begin{split}
        \langle Q^{2} \rangle &= \langle n | Q^{2} | n \rangle = \frac{\hbar}{2 \omega} \langle n | \left( a + a^{\dagger} \right)^{2} | n \rangle \\
        &= \frac{\hbar}{2 \omega} \langle n | a^{2} + \left( a^{\dagger} \right)^{2} + 2 a^{\dagger} a + 1 | n \rangle \\
        &= \frac{\hbar}{2 \omega} \left( 2n + 1 \right)
    \end{split}
\end{equation}
which shows that every additional phonon provides a contribution of $\frac{\hbar}{\omega}$. As such, we use a value of $\langle Q^{2} \rangle = \frac{\hbar}{\omega}$ to compute the polarization associated with a single phonon per unit cell with Eq. (4). For the magnetization, we use Eq. (5).

\subsection*{(Multi)ferron properties}

An overview of the single-phonon per unit cell properties can be found in Table~\ref{tab:single_phonon_multiferron} for the $A_1$ and $E$ modes, respectively. Among the $A_1$ modes, only the lowest (\SI{7.16}{\THz}) and highest frequency (\SI{18.24}{\THz}) modes exhibit a substantial polarization for a single phonon per unit cell of \SI{0.86}{\micro\coulomb\per\cm\squared} and \SI{0.50}{\micro\coulomb\per\cm\squared}, respectively. In both cases, the polarization opposes the ferroelectric polarization, reducing the net polarization (Fig. 1a). The relative change from a single-phonon excitation of these modes $\Delta \mathbf{P} = | \overline{\mathbf{P}} | / | \mathbf{P}_{0} | \cdot 100 \%$, with $\mathbf{P}_{0} = \SI{79.9}{\micro\coulomb\per\cm\squared}$, is about $1.1\%$ for the \SI{7.16}{\THz} mode and $0.6 \%$ for the \SI{18.24}{\THz} mode.

Turning to the degenerate $E$ modes, their in-plane character leads to a polarization oriented perpendicular to the ferroelectric axis (Fig. 1b), tilting the net polarization. As with the $A_1$ modes, the magnitude of this polarization changes depending on the mode, with the largest found for the \SI{10.54}{\THz} and \SI{17.04}{\THz} modes. The angle with which the polarization of LiNbO$_{3}$ is tilted, can be estimated through $\phi = \arctan \left( | \overline{\mathbf{P}} | / | \mathbf{P}_{0} | \right)$. For the \SI{10.54}{\THz} mode, which has the largest in-plane polarization, it is tilted by $\phi = \SI{0.18}{\degree}$.

\begin{table*}[htbp!]
\caption{\label{tab:single_phonon_multiferron}Calculated phonon frequencies $\nu_{0}$, mode effective charges $| \mathbf{Z} |$, and net single-phonon per unit cell polarization $| \overline{\mathbf{P}} |$ of the $A_1$ and $E$ modes in LiNbO$_{3}$. Single-phonon per unit cell magnetization $| \overline{\mathbf{M}} |$ of circularly polarized $E$ modes in LiNbO$_{3}$.}
\begin{ruledtabular}
    \begin{tabular}{cccccc}
    Symmetry & Mode & $\nu^{\mathrm{DFT}}_{0}$ (\SI{}{\THz}) & $| \mathbf{Z} |$ (\SI{}{\elementarycharge\per{\sqrt{\amu}}}) & $| \overline{\mathbf{P}} |$ (\SI{}{\micro\coulomb\per\cm\squared}) & $| \overline{\mathbf{M}} |$ ($\mu_{\mathrm{N}} / V_{\mathrm{C}}$) \\
    \colrule
    $A_1$               & 9         &  7.16     & 1.512     & 0.855     & -         \\
                        & 12        &  8.05     & 0.118     & 0.048     & -         \\
                        & 16        & 10.08     & 0.287     & 0.021     & -         \\
                        & 27        & 18.24     & 1.629     & 0.503     & -         \\
    \colrule
    $E$                 &  4,  5    &  4.32     & 1.020     & 0.148     & 0.100     \\
                        &  7,  8    &  6.40     & 0.295     & 0.017     & 0.113     \\
                        & 10, 11    &  7.61     & 0.946     & 0.135     & 0.063     \\
                        & 14, 15    &  9.44     & 0.704     & 0.047     & 0.097     \\
                        & 17, 18    & 10.54     & 0.790     & 0.248     & 0.129     \\
                        & 19, 20    & 10.88     & 0.158     & 0.010     & 0.119     \\
                        & 22, 23    & 12.35     & 0.256     & 0.006     & 0.092     \\
                        & 25, 26    & 17.04     & 1.647     & 0.222     & 0.043     \\
                        & 28, 29    & 19.92     & 0.277     & 0.001     & 0.013     \\
    \end{tabular}
\end{ruledtabular}
\end{table*}

\clearpage

\section{Laser pulses}

\subsection*{Functional form}

To model a laser pulse and its interaction with a material, we describe the time-dependent electric field generated by the pulse. The electric field is typically represented as a vector with components that vary in time. Here, we describe a laser pulse propagating along the $z$-axis by
\begin{equation}
    \mathbf{E}(t) = \frac{\tilde{E}_0}{\sqrt{2}} \exp \left \{ -\frac{(t - t_0)^2}{2 \left( \tau / \sqrt{8 \ln 2} \right)^2} \right \}
    \begin{bmatrix}
        \cos(\omega_0 t) \\
        \cos(\omega_0 t + \phi) \\
        0
    \end{bmatrix},
\end{equation}
where $\tau$ is the full width at half maximum (FWHM) of the pulse envelope, $\omega_0$ is the carrier angular frequency, and $\phi$ is the relative phase between the $x$- and $y$-components. The prefactor
\begin{equation}
    \tilde{E}_0 = \frac{2}{1 + \sqrt{\epsilon_{\infty}}} \, E_{0}
\end{equation}
represents the shielded peak electric field amplitude, reduced by the dielectric screening inside the material with the static dielectric constant $\epsilon_{\infty} = 5.72$ from DFT. The polarization state of the laser pulse is controlled by $\phi$: linear for $\phi = 0$, circular for $\phi = \pm \pi / 2$, and elliptical for intermediate values $0 < | \phi | < \pi / 2 $.

To explore the dynamics of different pairs of modes under comparable excitation conditions, we scale the fluence of the laser pulse with the energy of the modes ($\hbar \omega$). Following Ref.~\cite{juraschekOrbitalMagneticMoments2019}, we maintain a constant number of cycles in the laser pulse by fixing the ratio $\tau \omega_{0} = 5 \times 2\pi$. Additionally, we scale the electric field intensity quadratically with the laser frequency, using \SI{15}{\mega\volt\per\cm} at $\omega_{0}/(2\pi) = \SI{20}{\THz}$ as a reference point. As a result of this, we excite modes 4 and 5 with $E_{0} = \SI{0.70}{\mega\volt\per\cm}$ and $\tau = \SI{1.16}{\ps}$ and modes 17 and 18 with $E_{0} = \SI{4.16}{\mega\volt\per\cm}$ and $\tau = \SI{0.47}{\ps}$, both of which are experimentally feasible~\cite{liuGenerationNarrowbandHighintensity2017, liuPumpFrequencyResonances2020}.

\subsection*{Laser energy}

To ensure that the laser pulse delivers the same total energy to the material regardless of its polarization state, we verify that the cycle-averaged intensity is independent of $\phi$. This is done by computing the time-averaged amplitude of the electric field, without taking the pulse envelope into account.
\begin{enumerate}
    \item For a linearly polarized laser pulse ($\phi = 0$), the $x$- and $y$-components of the electric field are identical, i.e. $E_{x} \left( t \right) = E_{y} \left( t \right) = \frac{\tilde{E}_{0}}{\sqrt{2}} \cos \left( \omega_{0} t \right)$. The resulting instantaneous intensity is $| \mathbf{E} (t) |^2 = \tilde{E}_{0}^{2} \cos^{2} ( \omega_{0} t )$. Averaging over a complete cycle of the laser pulse yields $\langle |\mathbf{E}(t)|^2 \rangle = \tilde{E}_0^2 \langle \cos^2(\omega_0 t) \rangle = \frac{\tilde{E}_0^2}{2}$.
    
    \item Circular excitations ($\phi = \pm \pi / 2$) have out-of-phase $x$- and $y$-components for the electric field, for instance $E_{x} \left( t \right) = \frac{\tilde{E}_{0}}{\sqrt{2}} \cos \left( \omega_{0} t \right)$ and $E_{y} \left( t \right) = \frac{\tilde{E}_{0}}{\sqrt{2}} \sin \left( \omega_{0} t \right)$. The resulting instantaneous intensity is constant in time: $| \mathbf{E} \left( t \right) |^{2} = \frac{\tilde{E}^{2}_{0}}{2} \left[ \cos^{2} \left( \omega_{0} \right) + \sin^{2} \left( \omega_{0} t \right) \right] = \frac{\tilde{E}^{2}_{0}}{2}$.
    
    \item Elliptical excitations ($0 < | \phi | < \pi / 2 $) have $x$- and $y$-components for the electric field of $E_{x} \left( t \right) = \frac{\tilde{E}_{0}}{\sqrt{2}} \cos \left( \omega_{0} t \right)$ and $E_{y} \left( t \right) = \frac{\tilde{E}_{0}}{\sqrt{2}} \cos \left( \omega_{0} t + \phi \right)$, yielding an instantaneous intensity that can be rewritten using trigonometric identities to: $| \mathbf{E} \left( t \right) |^{2} = \frac{\tilde{E}_{0}^{2}}{2} \left[ \cos^{2} \left( \omega_{0} t \right) + \cos^{2} \left( \omega_{0} t + \phi \right) \right] = \frac{\tilde{E}_{0}^{2}}{4} \left[ 2 + \cos( 2 \omega_{0} t ) - \cos( 2 \omega_{0} t + 2 \phi) \right]$. Over a full cycle, it averages to $\langle | \mathbf{E} ( t ) |^{2} \rangle = \frac{\tilde{E}_{0}^{2}}{2}$.
\end{enumerate}

\begin{figure*}[htbp!]
    \includegraphics{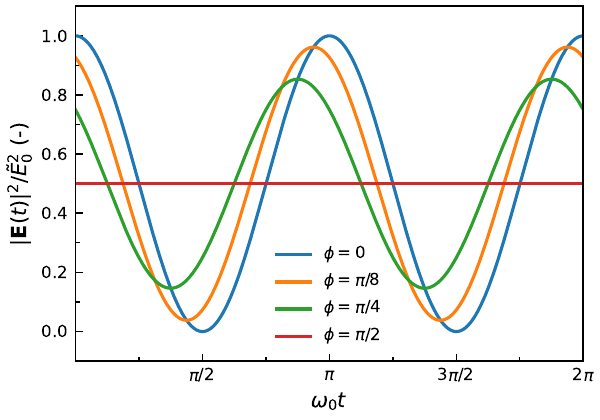}
    \caption{Squared electric field amplitude $| \mathbf{E} ( t ) |^{2} / \tilde{E}^{2}_{0}$ for laser pulses without an envelope with a polarization varying from linear ($\phi = 0$) to circular ($\phi = \pi/2$).}
    \label{fig-si:laser_intensities}
\end{figure*}

As shown above, all laser pulses, irrespective of their polarization, result in an identical cycle-averaged intensity. To demonstrate this, we show the electric field amplitude over a complete cycle for various laser polarization, i.e. $\phi = 0$, $\pi / 8$, $\pi / 4$, and $\pi / 2$ in Fig.~\ref{fig-si:laser_intensities}. Only the circularly polarized laser pulse has a constant electric field over time of $\frac{\tilde{E}_{0}^{2}}{2}$, for all other polarizations the electric field oscillates around the average value of $\frac{\tilde{E}_{0}^{2}}{2}$.

\subsection*{Lindemann criterion}

To assess whether the excitation with the laser pulse is in a reasonable range, we evaluate the Lindemann criterion for the induced atomic displacements~\cite{lindemannUberBerechnungMolekularer1910, sokolowski-tintenFemtosecondXrayMeasurement2003}. The criterion states that a crystal melts when the root mean square displacement $d_{\mathrm{rmsd}}$ of the atoms is $10\%-20\%$ of the interatomic distance in the crystal $d_0$. Using the laser pulse with  the $E$ modes at \SI{4.32}{\THz}, with $E_{0} = \SI{0.70}{\mega\volt\per\cm}$ and $\tau = \SI{1.16}{\ps}$, we find a value of $d_{\mathrm{rmsd}} / d_{0} = 2.8\%$ for the largest atomic displacement as seen for oxygen (O) species. The excitation is well below the range for which instabilities in the lattice are expected, predicting the crystal to remain intact.

\clearpage

\section{Polarization and magnetization dynamics}

\subsection*{High-frequency modes}

\begin{figure*}[htbp!]
    \includegraphics{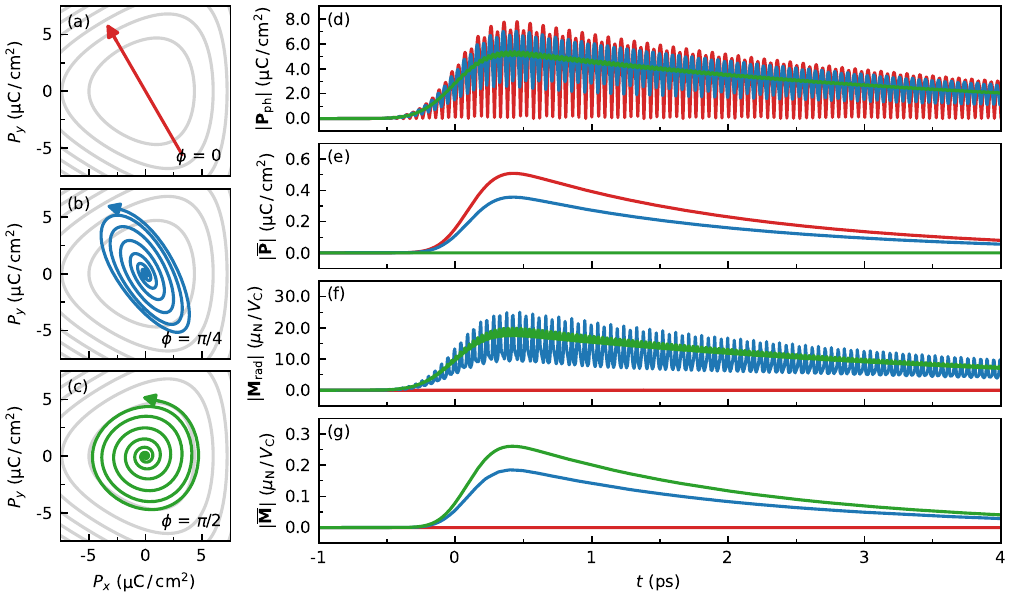}
    \caption{(a-c) Polarization dynamics driven by linearly ($\phi = 0$), elliptically ($\phi = \pi / 4$), and circularly ($\phi = \pi / 2$) polarized IR pulses in the time window $-1.0$ to $0.25$~\SI{}{\ps}, simulated with damping constant $\kappa = \omega / (20 \cdot 2 \pi)$}. Equipotential lines representing the symmetry of the polarization are shown in gray. (d-e) Time evolution of the radial polarization $| \mathbf{P}_{\mathrm{ph}} |$ and net polarization $| \overline{\mathbf{P}} |$. (f-g) Corresponding radial magnetization $|\mathbf{M}_{\mathrm{rad}}|$ and net magnetization $|\overline{\mathbf{M}}|$.
    \label{fig-si:high-frequency_modes}
\end{figure*}

To demonstrate that the polarization and magnetization dynamics shown in the main text are not unique to modes 4 and 5 at \SI{4.32}{\THz}, we also present the dynamics of modes 17 and 18 in Fig.~\ref{fig-si:high-frequency_modes}. These modes have a frequency of \SI{10.54}{\THz}, as evidenced by the more rapid fluctuations in transient polarization and magnetization signals (Fig.~\ref{fig-si:high-frequency_modes}d,f).

\clearpage

\subsection*{Tunability of multiferrons}

\begin{figure*}[htbp!]
    \includegraphics{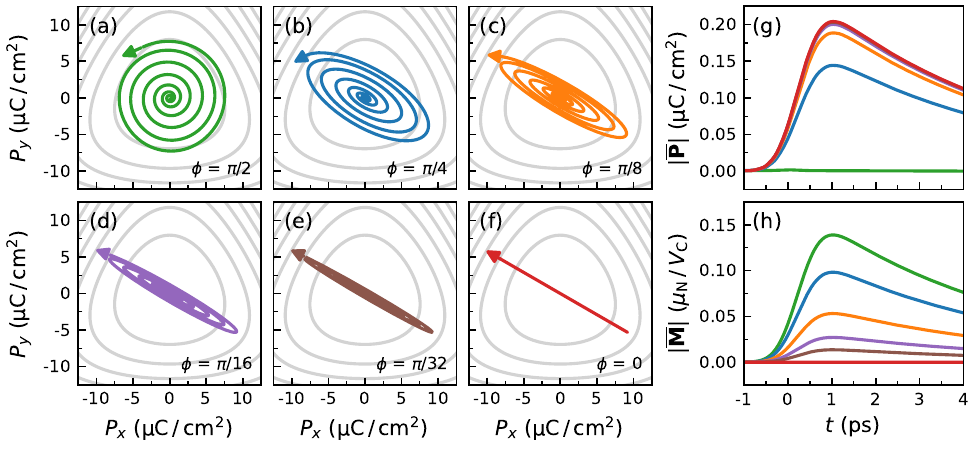}
    \caption{Effects of ellipticity of the laser pulse on the  of degenerate $E$ modes at \SI{4.32}{\THz} in \ch{LiNbO3}. (a-f) Polarization dynamics driven by IR pulses with various character in the time window from $-1.0$ to $0.5$~\SI{}{\ps}. Equipotential lines representing the symmetry of the polarization are shown in gray. (g-h) Time evolution of the net polarization $| \overline{\mathbf{P}} |$ and net magnetization $| \overline{\mathbf{M}} |$.}
    \label{fig-si:ellipticity_tuning}
\end{figure*}

The relative magnitudes of the dipole moment and magnetic moment that emerge as a result of a laser excitation can be tuned through the polarization of the pulse, as shown in Fig.~\ref{fig-si:ellipticity_tuning}. A circularly polarized ($\phi = \pi / 2$) pulse yields and excitation without a net dipole moment, but with a net magnetic moment. By making the pulse elliptical ($\phi = \pi / 4$), the magnetic moment of the excitation decreases, while simultaneously creating a net dipole moment. Further increasing the ellipticity of the laser pulse ($\phi = \pi / 8$, $\pi / 16$, and $\pi / 32$) results in a larger net dipole moment, but a smaller magnetic moment. For a linear excitation ($\phi = 0$), the net dipole moment is largest, whereas the magnetic moment has vanished.

\clearpage

\section{Multipole moments}

\subsection*{Quadrupole moments}

\begin{figure*}[htbp!]
    \includegraphics{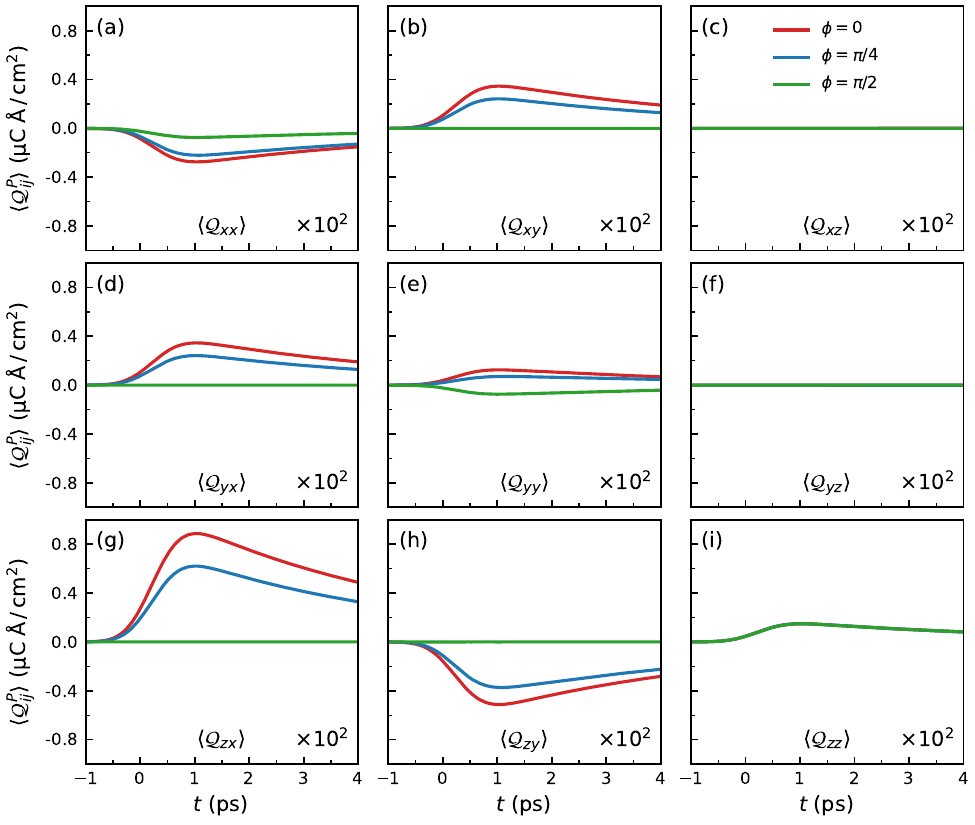}
    \caption{Time-average quadrupole tensor of the radial polarization $| \mathbf{P}_{\mathrm{ph}} |$. (a-i) All tensor components $\langle \mathcal{Q}_{ij} \rangle$ with $i,j \in \{ x, y, z \}$ are shown for linearly ($\phi = 0$), elliptically ($\phi = \pi / 4$), and circularly ($\phi = \pi / 2$) polarized pulses.}
    \label{fig-si:electric_quadrupole}
\end{figure*}

\begin{figure*}[htbp!]
    \includegraphics{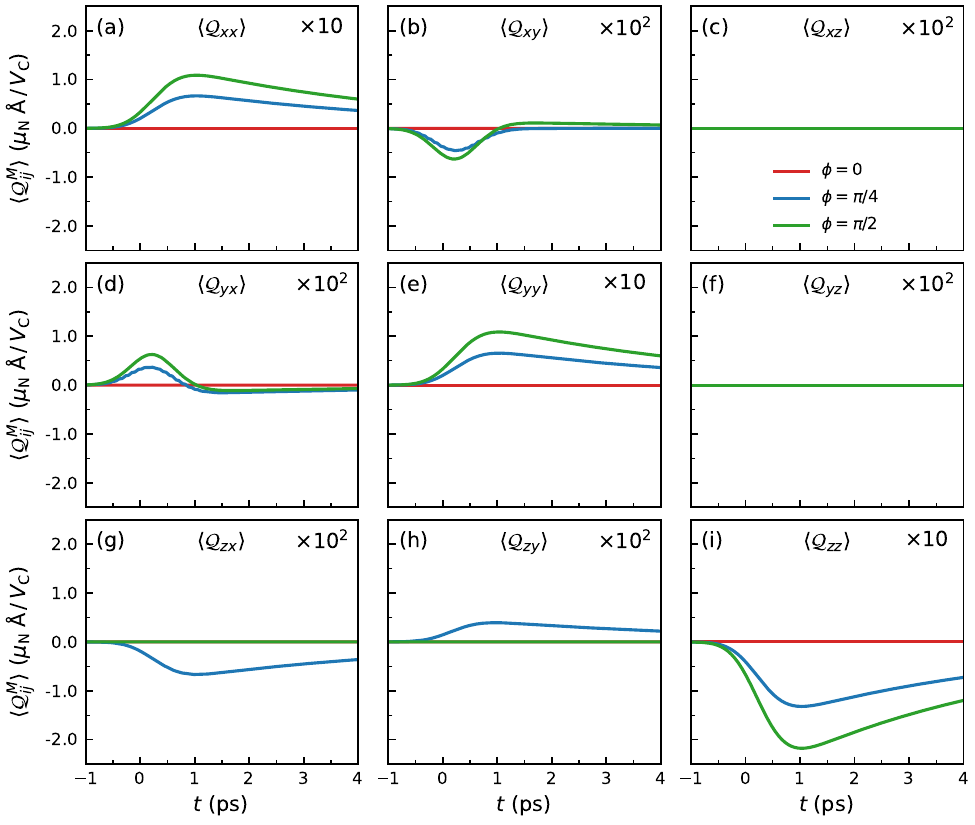}
    \caption{Time-average quadrupole tensor of the radial magnetization $| \mathbf{M}_{\mathrm{rad}} |$. (a-i) All tensor components $\langle \mathcal{Q}_{ij} \rangle$ with $i,j \in \{ x, y, z \}$ are shown for linearly ($\phi = 0$), elliptically ($\phi = \pi / 4$), and circularly ($\phi = \pi / 2$) polarized pulses.}
    \label{fig-si:magnetic_quadrupole}
\end{figure*}

For completeness, we here show all quadrupole tensor components for the radial polarization $\mathbf{P}_{\mathrm{ph}}$ and radial magnetization $\mathbf{M}_{\mathrm{rad}}$. We used Eq. (10) and Eq. (11) to determine the quadrupole contributions shown in Fig.~\ref{fig-si:electric_quadrupole} and Fig.~\ref{fig-si:magnetic_quadrupole}.

\begin{figure*}[htbp!]
    \includegraphics{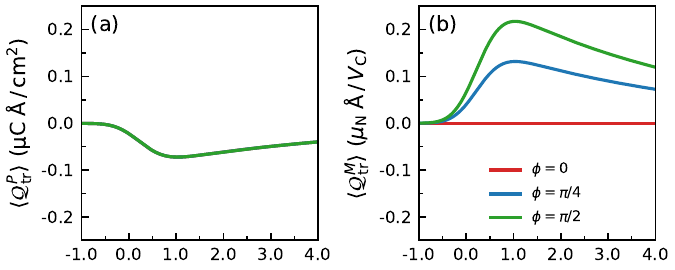}
    \caption{Time-average trace of the quadrupole tensors of (a) radial polarization $\mathbf{P}_{\mathrm{ph}}$ and (b) $\mathbf{M}_{\mathrm{rad}}$ for linearly ($\phi = 0$), elliptically ($\phi = \pi / 4$), and circularly ($\phi = \pi / 2$) polarized pulses.}
    \label{fig-si:quadrupole_trace}
\end{figure*}

In Fig.~\ref{fig-si:quadrupole_trace} we show the trace of the quadrupole tensors $\langle \mathcal{Q}_{\mathrm{tr}} \rangle$ that we define as
\begin{align}
    \langle \mathcal{Q}^{P}_{\mathrm{tr}} \rangle = \frac{1}{3} \sum_{n} \sum_{i} \langle u_{n,i} P_{\mathrm{ph},n,i} \rangle \\
    \langle \mathcal{Q}^{M}_{\mathrm{tr}} \rangle = \frac{1}{3} \sum_{n} \sum_{i} \langle u_{n,i} M_{\mathrm{rad},n,i} \rangle
\end{align}
with $n$ summing over the atoms, and $i$ over the Cartesian directions $x$, $y$, and $z$.

\subsection*{Octupole moments}

\begin{figure*}[htbp!]
    \includegraphics{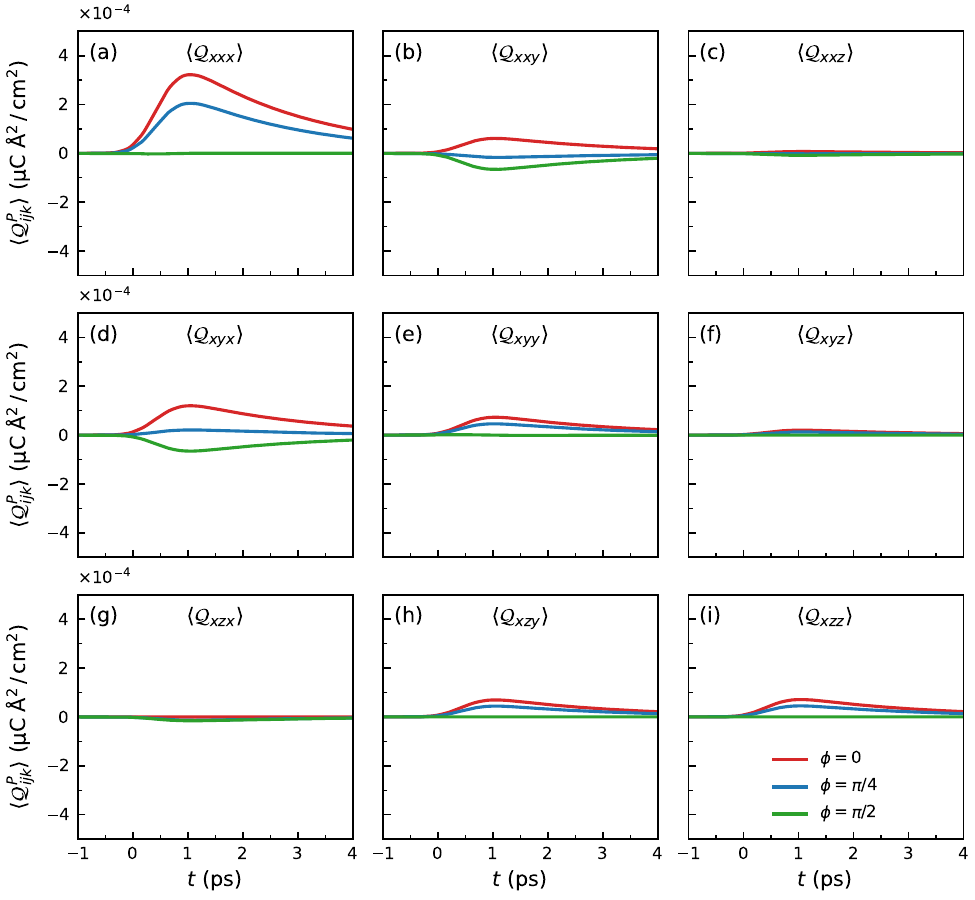}
    \caption{Time-average octupole tensor of the radial polarization $| \mathbf{P}_{\mathrm{ph}} |$. (a-i) All tensor components $\langle \mathcal{O}_{xjk} \rangle$ with $j,k \in \{ x, y, z \}$ are shown for linearly ($\phi = 0$), elliptically ($\phi = \pi / 4$), and circularly ($\phi = \pi / 2$) polarized pulses.}
    \label{fig-si:electric_octupole_x}
\end{figure*}

\begin{figure*}[htbp!]
    \includegraphics{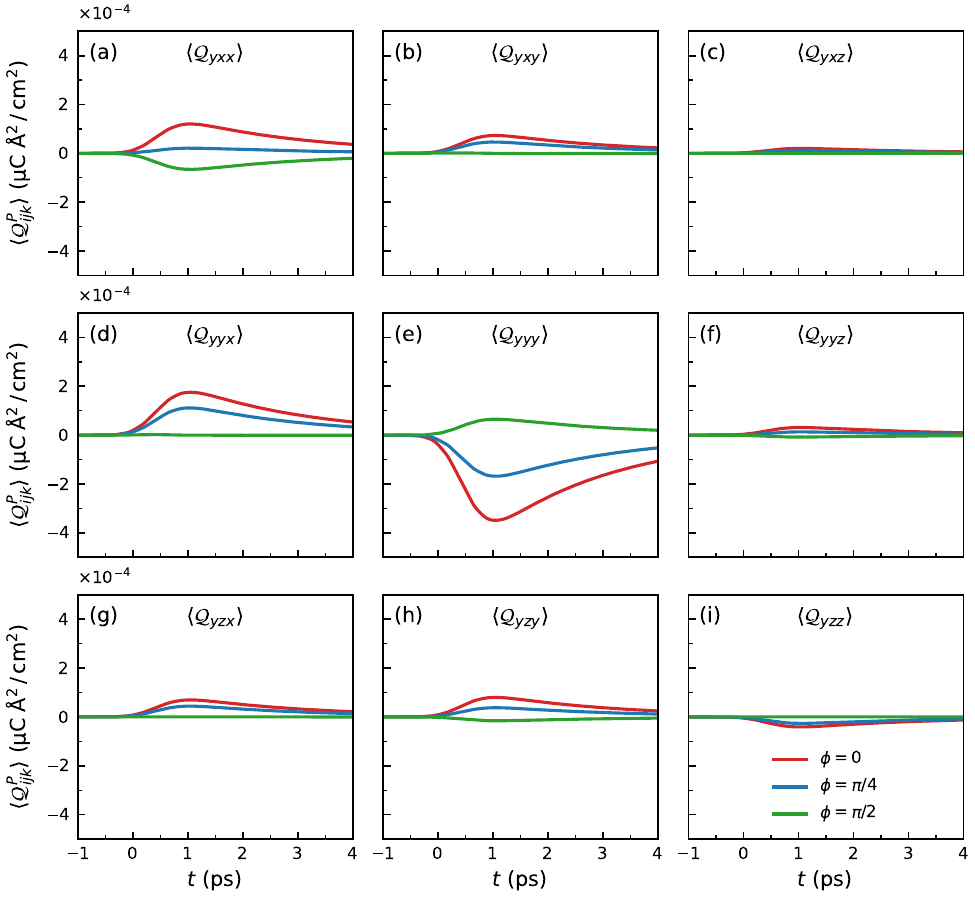}
    \caption{Time-average octupole tensor of the radial polarization $| \mathbf{P}_{\mathrm{ph}} |$. (a-i) All tensor components $\langle \mathcal{O}_{yjk} \rangle$ with $j,k \in \{ x, y, z \}$ are shown for linearly ($\phi = 0$), elliptically ($\phi = \pi / 4$), and circularly ($\phi = \pi / 2$) polarized pulses.}
    \label{fig-si:electric_octupole_y}
\end{figure*}

\begin{figure*}[htbp!]
    \includegraphics{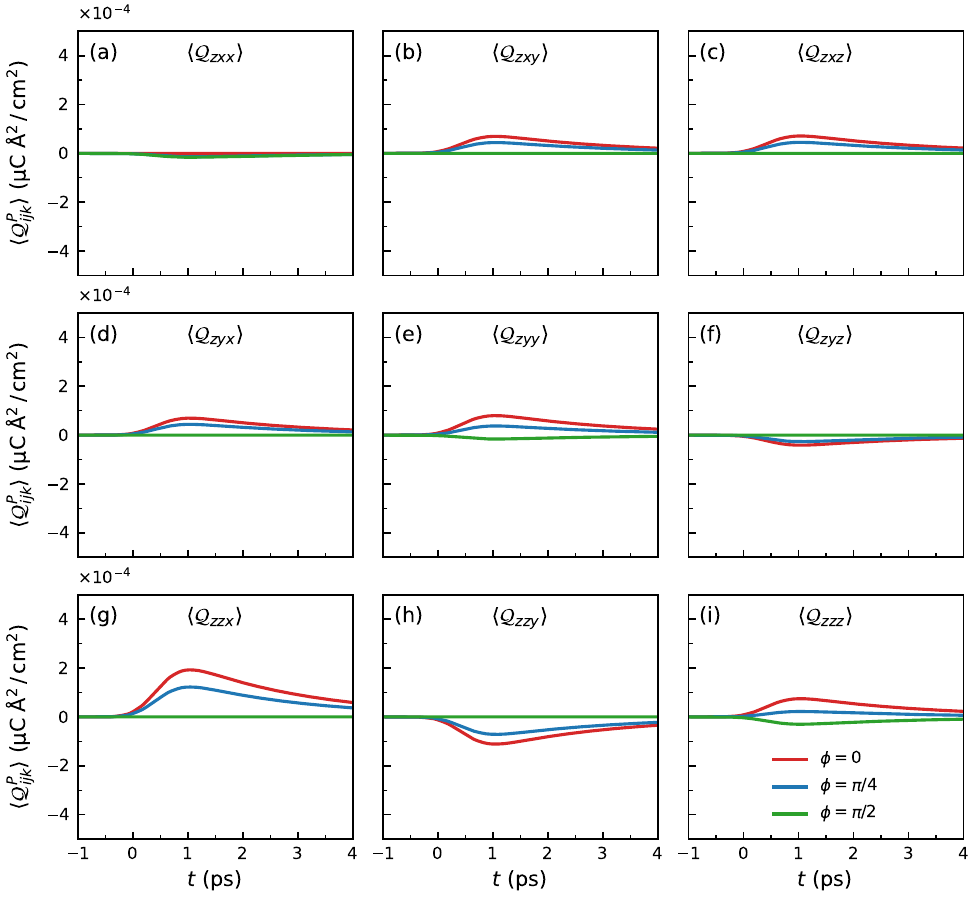}
    \caption{Time-average octupole tensor of the radial polarization $| \mathbf{P}_{\mathrm{ph}} |$. (a-i) All tensor components $\langle \mathcal{O}_{zjk} \rangle$ with $j,k \in \{ x, y, z \}$ are shown for linearly ($\phi = 0$), elliptically ($\phi = \pi / 4$), and circularly ($\phi = \pi / 2$) polarized pulses.}
    \label{fig-si:electric_octupole_z}
\end{figure*}

\begin{figure*}[htbp!]
    \includegraphics{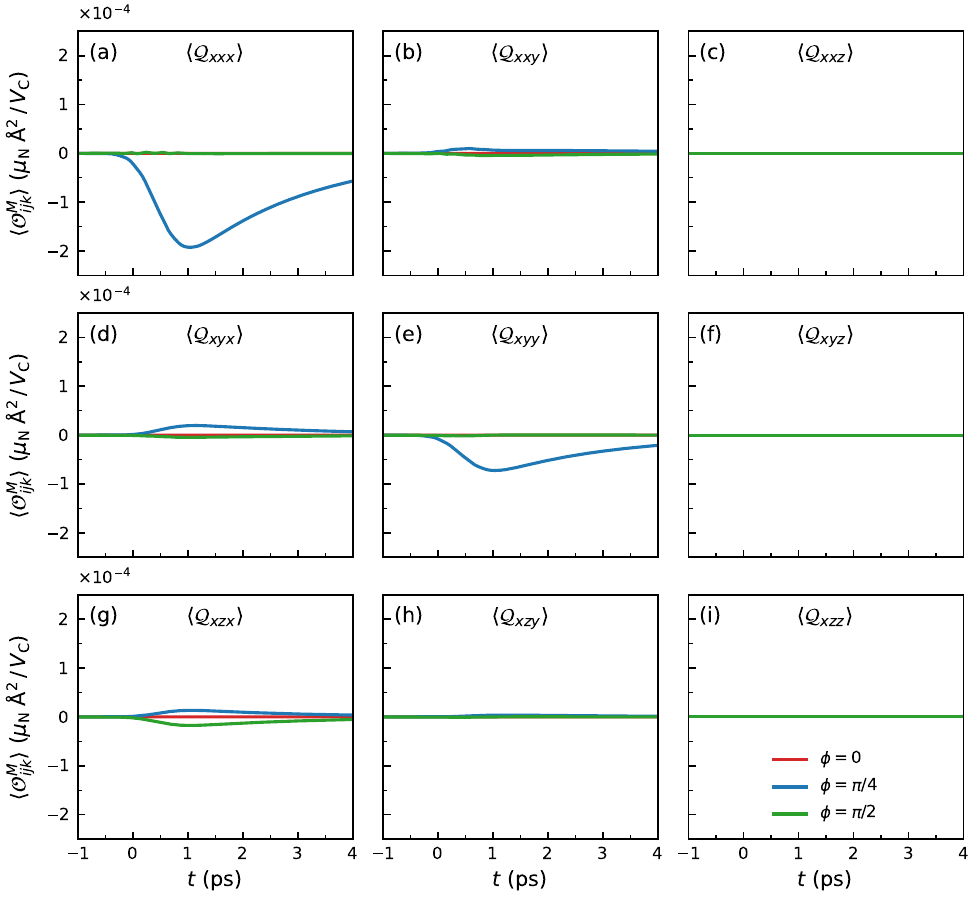}
    \caption{Time-average octupole tensor of the radial magnetization $| \mathbf{M}_{\mathrm{rad}} |$. (a-i) All tensor components $\langle \mathcal{O}_{xjk} \rangle$ with $j,k \in \{ x, y, z \}$ are shown for linearly ($\phi = 0$), elliptically ($\phi = \pi / 4$), and circularly ($\phi = \pi / 2$) polarized pulses.}
    \label{fig-si:magnetic_octupole_x}
\end{figure*}

\begin{figure*}[htbp!]
    \includegraphics{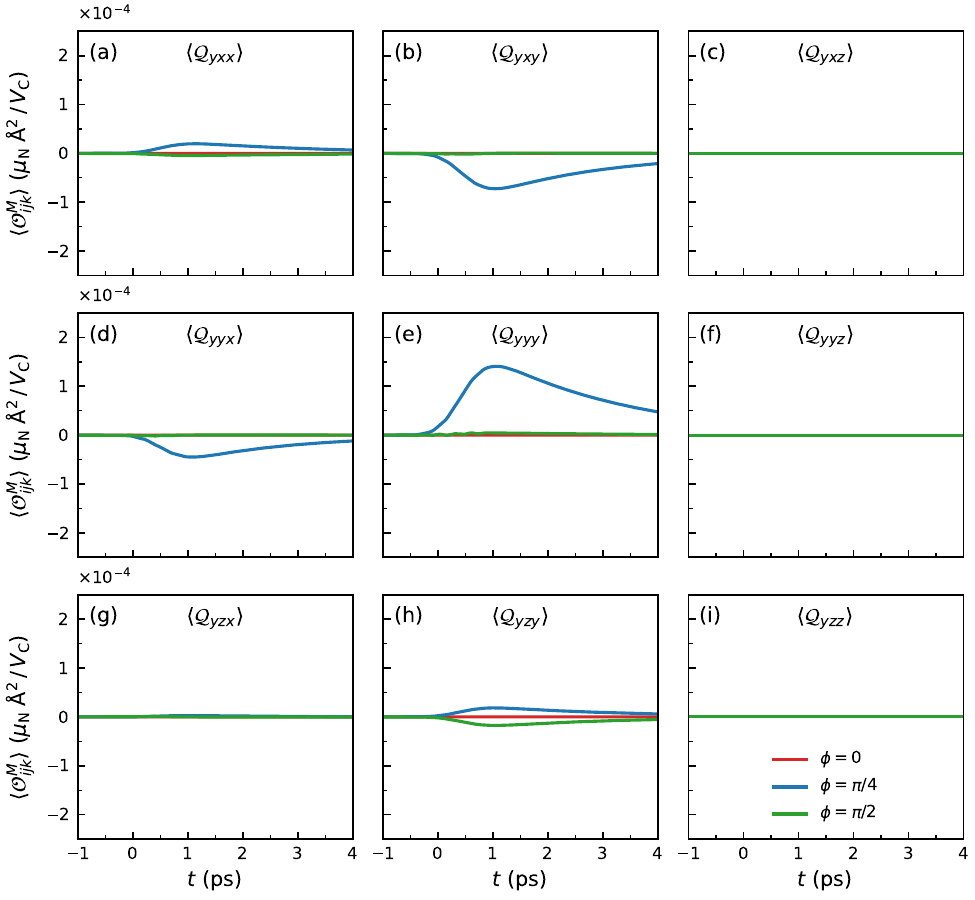}
    \caption{Time-average octupole tensor of the radial magnetization $| \mathbf{M}_{\mathrm{rad}} |$. (a-i) All tensor components $\langle \mathcal{O}_{yjk} \rangle$ with $j,k \in \{ x, y, z \}$ are shown for linearly ($\phi = 0$), elliptically ($\phi = \pi / 4$), and circularly ($\phi = \pi / 2$) polarized pulses.}
    \label{fig-si:magnetic_octupole_y}
\end{figure*}

\begin{figure*}[htbp!]
    \includegraphics{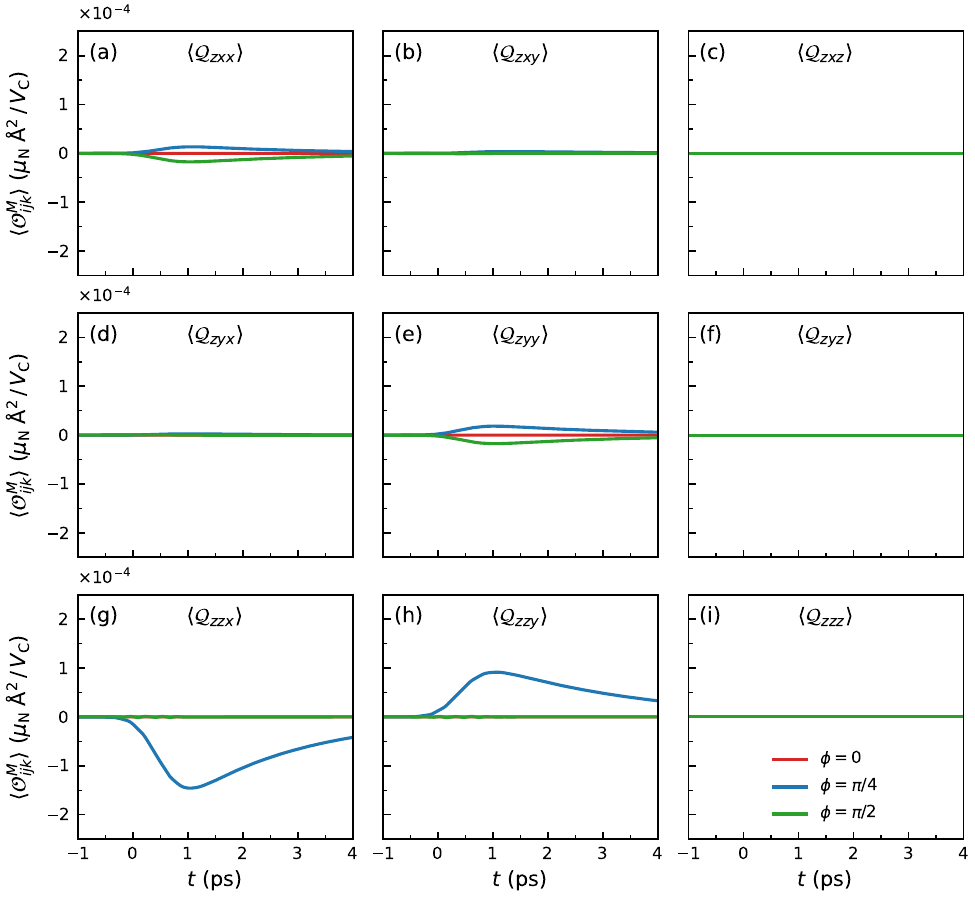}
    \caption{Time-average octupole tensor of the radial magnetization $| \mathbf{M}_{\mathrm{rad}} |$. (a-i) All tensor components $\langle \mathcal{O}_{zjk} \rangle$ with $j,k \in \{ x, y, z \}$ are shown for linearly ($\phi = 0$), elliptically ($\phi = \pi / 4$), and circularly ($\phi = \pi / 2$) polarized pulses.}
    \label{fig-si:magnetic_octupole_z}
\end{figure*}

Going beyond the quadrupole contributions, we can determine the octupole tensor components $\mathcal{O}_{ijk}$ of the radial polarization $\mathbf{P}_{\mathrm{ph}}$ and radial magnetization $\mathbf{M}_{\mathrm{rad}}$. The atomic contribution of atom $n$ to the octupole tensor can be computed as
\begin{align}
    \label{eq:octupole_tensor}
    \langle \mathcal{O}^{P}_{n,ijk} \rangle &= \left \langle u_{n,i} \left( t \right) u_{n,j} \left( t \right) P_{\mathrm{ph},n,k} \left( t \right) \right \rangle \\
    \langle \mathcal{O}^{M}_{n,ijk} \rangle &= \left \langle u_{n,i} \left( t \right) u_{n,j} \left( t \right) M_{\mathrm{rad},n,k} \left( t \right) \right \rangle
\end{align}
with $\langle \cdots \rangle$ denoting an average over the phonon period and $i,j,k \in \{ x, y, z \}$ indicating the Cartesian components of the vectors. The total octupole tensor components then follow from a summation over all atomic contributions: $\langle \mathcal{O}_{ijk} \rangle = \sum_{n} \langle \mathcal{O}_{n,ijk} \rangle$. A complete overview of the octupole components of the radial polarization and radial magnetization can be found in Figs.~\ref{fig-si:electric_octupole_x}-\ref{fig-si:electric_octupole_z} and Figs.~\ref{fig-si:magnetic_octupole_x}-\ref{fig-si:magnetic_octupole_z}, respectively.

\clearpage

\bibliography{ref}